\documentclass[12pt,a4paper]{article}
\usepackage{a4wide}
\usepackage{latexsym}
\usepackage{epsf}
\usepackage{amssymb}
\makeatletter
\@addtoreset{equation}{section}
\makeatother

\begin{document}

\begin{flushright}
\small
IFT-UAM/CSIC-03-53\\
{\bf hep-th/0401005}\\
January $2$nd, $2004$
\normalsize
\end{flushright}

\begin{center}


\vspace{2cm}

{\Large {\bf G\"odel Spacetimes, Abelian Instantons, the Graviphoton
    Background and other Flacuum Solutions}}

\vspace{2cm}


{\bf\large Patrick Meessen\footnote{E-mail: 
{\tt Patrick.Meessen@uam.es}} 
and Tom\'as Ort\'{\i}n}${}^{\spadesuit}$
\footnote{E-mail: {\tt Tomas.Ortin@cern.ch}}
\vskip 0.4truecm

\vspace{.7cm}

{\it Instituto de F\'{\i}sica Te\'orica, C-XVI,
Universidad Aut\'onoma de Madrid \\
E-28049-Madrid, Spain}

\vspace{2cm}


{\bf Abstract}

\end{center}

\begin{quotation}

\small

We study the relations between all the vacua of Lorentzian and
Euclidean $d=4,5,6$ SUGRAs with 8 supercharges, finding a new limiting
procedure that takes us from the over-rotating near-horizon BMPV black
hole to the G\"odel spacetime.  The timelike compactification of the
maximally supersymmetric G\"odel solution of $N=1,d=5$ SUGRA gives a
maximally supersymmetric solution of pure Euclidean $N=2,d=4$ with
flat space but non-trivial anti-selfdual vector field flux
(\textit{flacuum}) that, on the one hand, can be interpreted as an
$U(1)$ instanton on $T^{4}$ and that, on the other hand, coincides with
the graviphoton background shown by Berkovits and Seiberg to produce
the C-deformation introduced recently by Ooguri and Vafa.  We
construct flacuum solutions in other theories such as Euclidean
type~IIA supergravity.

\end{quotation}

\newpage

\pagestyle{plain}


\tableofcontents


\section{Introduction}

There has been much progress recently in the classification of
supersymmetric vacua in $d=4,5,6$ Lorentzian supergravities (SUGRAs)
with eight supercharges
\cite{Gauntlett:2002nw,Gutowski:2003rg,Chamseddine:2003yy,Caldarelli:2003pb}
and in the study of their relations via dimensional reduction and
oxidation \cite{Lozano-Tellechea:2002pn,Fiol:2003yq}. The study of
these theories and their supersymmetric vacua is interesting not just
as warm-up exercise for the more complicated cases of interest in
string theory but also because these theories arise in dimensional
reductions of 10-dimensional superstring effective theories and the
solutions found can be uplifted to full 11- or 10-dimensional M/string
theory vacua still preserving a large amount of supersymmetry.

It is in this way that the supersymmetric G\"odel spacetime solutions
of M/string theory have been found: as a maximally supersymmetric
vacuum of $N=1,d=5$ SUGRA which could be uplifted to 10 and 11
dimensions \cite{Gutowski:2003rg}. Many higher- \cite{Harmark:2003ud}
and also lower-dimensional \cite{Caldarelli:2003wh} examples of
supersymmetric supergravity solutions with a spacetime geometry
similar to the original 4-dimensional cosmological solution by
K.~G\"odel \cite{Godel:ga} (henceforth referred to as G\"odel
spacetimes) have been found after it was observed that G\"odel
spacetimes are T-duals of H$pp$-wave spacetimes\footnote{G\"odel
  spacetimes had previously been found as solutions of compactified
  string theory in \cite{Horowitz:1994rf,Russo:1994cv}. More recent
  solutions of supergravities in various dimensions with G\"odel
  spacetimes have been constructed in \cite{Gurses:2003}.}.  This fact
has raised many questions about the consistency of string theory in
such spacetimes which, being supersymmetric vacua on which (for
instance) black holes can be placed \cite{Gimon:2003ms,Brecher:2003wq}
and on which strings may be quantized
\cite{Harmark:2003ud,Blau:2003ia,Brace:2003st}, have generically
closed timelike curves (CTCs). This is one of the various pathologies
of General Relativity that supersymmetry or string theory are supposed
to cure, forbid or explain in a consistent way and a great deal of
interesting work has been done in this direction
\cite{Fiol:2003yq,Boyda:2002ba,Drukker:2003sc,Hikida:2003yd,Brecher:2003rv}.

The supersymmetric vacua of M/string theory provide an interesting
arena on which to study the general problems of quantization on curved
spacetimes, due to the high degree of (super) symmetry they have,
which, through the definition of conserved (super) charges and
symmetry (super) algebra, gives a good control and understanding of
the kinematics of the field theories defined on them. The fact that
the aDS/CFT correspondence ``works'' is largely owed to the
coincidence of the symmetry superalgebras of the theories involved.

The symmetry (super) algebras of the M/string theory vacua (and of the
field theories defined on them) arise deformations of the symmetry
(super) algebra of the original theory of which the vacua are
classical solutions, which is in general the Poincar\'e or anti-De
Sitter (super) algebra. This suggests that quantization on curved
backgrounds can be, at least in some cases, equivalent to quantization
using a deformation of the standard Heisenberg algebra. In other
words: quantum field theories in non-trivial backgrounds can be
equivalent to certain non-commutative field theories. An example of
this more general relation has been provided recently by Berkovits and
Seiberg \cite{Seiberg:2003yz,Berkovits:2003kj} who have found a
background of Euclidean $N=2,d=4$ SUGRA whose symmetry superalgebra
corresponds to the C-deformation introduced recently by Ooguri and
Vafa \cite{Ooguri:2003qp}. We have rediscovered this background in our
study of maximally supersymmetric vacua of $d=4,5,6$ SUGRAs with 8
supercharges, in relation with the G\"odel solution discussed above,
as we are going to explain.

\begin{figure}[!ht]
\begin{center}
\leavevmode
\epsfxsize= 9cm
\epsffile{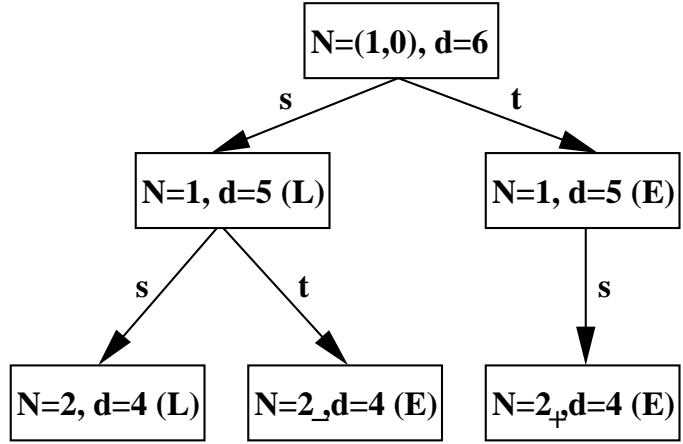}
\caption{Relations between the different Euclidean (\textbf{E}) 
  and Lorentzian (\textbf{L}) $d=4,5,6$ SUGRA theories with 8
  supercharges by timelike (\textbf{t}) or spacelike (\textbf{s})
  dimensional reductions. There is no Euclidean continuation of the
  Lorentzian $N=(1,0),d=6$ theory, due to the anti-selfduality of its
  3-form field strength, and there are two different Euclidean
  $N=2,d=4$ SUGRAs related by analytical continuation of the vector
  field.}
\label{fig:theories}
\end{center}
\end{figure}

Our goal in this paper is to complete our knowledge of the relations
between the maximally supersymmetric vacua of $d=4,5,6$ SUGRAs with 8
supercharges, extending our previous work
\cite{Lozano-Tellechea:2002pn} to Euclidean SUGRAs which arise in
timelike dimensional reductions (see Figure~\ref{fig:theories}). Our
main results are summarized in Figure~\ref{fig:d456vacua} and include
the identification of the above mentioned maximally supersymmetric
graviphoton background of Euclidean $N=2,d=4$ SUGRA that cannot be
obtained by Wick-rotating any known solution of the Lorentzian theory.
This solution can be obtained by timelike dimensional reduction of the
5-dimensional maximally supersymmetric G\"odel vacuum and has a flat
Euclidean space geometry in presence of a non-trivial anti-selfdual
2-form flux which has suggested the name \textit{flacuum} solution for
it.  Solutions of this kind are just as generic as those of G\"odel
type since they can be obtained by timelike compactifications of these
and we study some examples and, in more detail, the 4-dimensional one
which can also be interpreted as a $U(1)$ instanton over a $T^{4}$.

The existence of these solutions is due to the fact that the Euclidean
``energy-momentum'' tensor (the variation of the matter Lagrangian
with respect to the metric) is not positive- or negative-definite and
can vanish for non-trivial fields\footnote{The roles and definiteness
  of the energy and action are interchanged when we go from Lorentzian
  to Euclidean signature.}. This vanishing can be due to a
cancellation between ``energy-momentum'' tensors of different fields
(as in the case of the Einstein-frame D-instanton solution
\cite{Giddings:1989bq,Gibbons:1995vg}) or to a cancellation between
``electric'' and ``magnetic'' components of a single field. In the
case of a vector field in $d=4$ Euclidean space, the
``energy-momentum'' tensor vanishes whenever the field strength is
selfdual or anti-selfdual, as in the graviphoton background
\cite{Seiberg:2003yz,Berkovits:2003kj}.

A standard procedure to construct Euclidean theories is to
dimensionally reduce a Lorentzian theory in a timelike direction.
This procedure avoids the difficulties of the Wick rotation of
fermions and provides a mechanism to generate flacuum solutions from
higher-dimensional Lorentzian solutions with metrics of the form

\begin{equation}
d\hat{s}^{2}= (dt+\omega)^{2}-d\vec{x}_{d}^{2}\, ,
\end{equation}

\noindent
which is precisely the general form of the supersymmetric G\"odel
solutions studied in \cite{Harmark:2003ud}. This observation will
allow us to construct flacuum solutions in $d=10$ dimensions.

This paper is organized as follows: in Section~\ref{sec-dimred1} we
summarize the dimensional reduction of standard
(Lorentzian)\footnote{There is no Euclidean continuation of the
  $N=(1,0),\hat{d}=6$ theory because the anti-selfduality condition of
  the 3-form field strength cannot be consistently defined in
  Euclidean signature. The minimal Euclidean SUGRA in six dimensions
  has not yet been constructed. It would be interesting to construct
  this theory since it must be related by spacelike dimensional
  reduction to the same Euclidean $N=1,d=5$ SUGRA that we are going to
  obtain from timelike dimensional reduction.}  $N=(1,0),\hat{d}=6$
SUGRA in an arbitrary spacelike or timelike direction and its
truncation to pure Lorentzian or Euclidean $N=1,d=5$ SUGRA.  In
Section~\ref{sec-dimred2} we summarize the dimensional reduction of
the Lorentzian and Euclidean $N=1,\hat{d}=5$ SUGRAs obtained in the
previous section in an arbitrary spacelike or timelike direction and
its truncation to pure Lorentzian or Euclidean $N=2,d=4$ SUGRAs.  The
details of these reductions are explained in
Appendix~\ref{app-details}, but the all the main formulae necessary
for oxidation and reduction of solutions are contained in the first
two sections.  Observe that, as shown in Figure~\ref{fig:theories}, we
obtain two different Euclidean $N=2,d=4$ SUGRAs. These two theories
are related by an analytical continuation of the vector field.  In the
next two sections \ref{sec-vacuum5} and \ref{sec-vacuum4} we apply the
results of the first two to maximally supersymmetric vacua of
$N=(1,0),d=6$ SUGRA to find maximally supersymmetric vacua of pure
Euclidean and Lorentzian $N=1,d=5$ and $N=2,d=4$ SUGRA.  We only find
one not previously known from the reduction of the G\"odel solution,
which coincides with the graviphoton background of Seiberg and
Berkovits \cite{Seiberg:2003yz,Berkovits:2003kj}.  This solution, with
flat space and constant anti-selfdual field strength, has interesting
properties that we study in detail. In Section~\ref{sec-selfdual} we
study more general flacuum solutions in higher dimensions.
Section~\ref{sec-conclusions} contains our conclusions.

\section{Dimensional reduction of $N=(1,0),d=6$ SUGRA 
  in the timelike direction}
\label{sec-dimred1}

$N=(1,0),d=6$ SUGRA\footnote{Our conventions are essentially those of
  \cite{Lozano-Tellechea:2002pn} which are those of \cite{Nishino:dc}
  with some changes in the normalizations of the fields. In
  particular, we use mostly minus signature and Latin (resp.~Greek or
  underlined) letters for Lorentz (resp.~curved) indices. Further,
  $\hat{\gamma}_{7}=\hat{\gamma}_{0}\cdots \hat{\gamma}_{5}$,
  $\hat{\gamma}_{7}^{2}=+1$, $\hat{\epsilon}^{012345}=+1$,
  $\hat{\gamma}^{\hat{a}_{1}\cdots \hat{a}_{n}} =
  \frac{(-1)^{[n/2]}}{(6-n)!}  \hat{\epsilon}^{\hat{a}_{1}\cdots
    \hat{a}_{n}\hat{b}_{1}\cdots \hat{b}_{6-n}}
  \hat{\gamma}_{\hat{b}_{1}\cdots \hat{b}_{6-n}} \hat{\gamma}_{7}$.
  Positive and negative chiralities are defined by
  $\hat{\gamma}_{7}\hat{\psi}^{\pm}=\pm \hat{\psi}^{\pm}$. We use the
  notation $\not\! F_{a}\equiv \gamma^{b}F_{ba}$, $\not\!  F\equiv
  \gamma^{ab}F_{ab}$ and $\not\! H\equiv \gamma^{abc}H_{abc}$.}, the
minimal 6-dimensional SUGRA, consists of the metric
$\hat{e}^{\hat{a}}{}_{\hat{\mu}}$, 2-form field
$\hat{B}^{-}_{\hat{\mu}\hat{\nu}}$ with anti-self-dual field strength
$\hat{H}^{-}=3\partial\hat{B}^{-}$ and positive-chirality symplectic
Majorana-Weyl gravitino that can be described as a single, complex
Dirac spinor satisfying a Weyl condition $\hat{\psi}^{+}_{\hat{\mu}}$.
The bosonic equations of motion can be derived from the action

\begin{equation}
\label{eq:d6action}
\hat{S}=\int d^{6}\hat{x}\sqrt{|\hat{g}|}\, \left[\hat{R}
+{\textstyle\frac{1}{24}} (\hat{H}^{-})^{2}\right]\, ,  
\end{equation}

\noindent 
imposing afterwards the anti-self-duality constraint
${}^{\star}\hat{H}^{-}=-\hat{H}^{-}$. The gravitino supersymmetry
transformation rule is (for zero fermions)

\begin{equation}
\label{eq:d6susy}
\delta_{\hat{\epsilon}^{+}}\hat{\psi}^{+}_{\hat{a}} 
=\left(\hat{\nabla}_{\hat{a}}  
-{\textstyle\frac{1}{48\sqrt{2}}}\not\!\!\hat{H}^{-}\hat{\gamma}_{\hat{a}} 
\right)\hat{\epsilon}^{+}\, .
\end{equation}

We can perform the dimensional reduction\footnote{In this section we
  use hats for 6-dimensional objects and no hats for 5-dimensional
  objects. In general we will use hats for higher-dimensional
  ($\hat{d}$-dimensional) indices and no hats for
  $d=\hat{d}-1$-dimensional indices. The index corresponding to the
  dimension which is being reduced is $\sharp$ which will be $t$ in
  the timelike case or $z$ spacelike case.  Observe that, with our
  mostly minus convention we get a negative-definite $d$-dimensional
  metric in the timelike case
  $\eta_{ab}=\hat{\eta}_{ab}=-\delta_{ab}$.} of $N=(1,0),\hat{d}=6$
SUGRA in one dimension can be performed simultaneously for the cases
in which that dimension is timelike
($\hat{\eta}_{\sharp\sharp}\equiv\alpha_{\hat{d}}=+1$) or spacelike
($\hat{\eta}_{\sharp\sharp}\equiv\alpha_{\hat{d}}=-1$). The reduction
gives $N=1,d=5$ SUGRA (the minimal 5-dimensional SUGRA) (metric
$e^{a}{}_{\mu}$, {\it graviphoton} $V_{\mu}$ and symplectic-Majorana
gravitino that can be described as a single complex, unconstrained
Dirac spinor $\psi_{\mu}$ \cite{Cremmer:1980gs}) coupled to a vector
multiplet consisting of a gaugino $\lambda$ (the 6th component of the
6-dimensional gravitino, a real scalar (the KK one $k$) and a vector
field $W_{\mu}$. $V_{\mu}$ and $W_{\mu}$ are combinations of scalars,
the KK vector field $A_{\mu}$ and the vector field that comes from the
6-dimensional 2-form $B_{\mu}$. The identification of the right
combinations is made by imposing consistency of the truncation to pure
$N=1,d=5$ SUGRA\footnote{The details of this dimensional reduction and
  truncation are explained in Appendix~\ref{app-details}.}.

The result can be stated as follows: any solution of $N=(1,0),d=6$
SUGRA satisfying the truncation constraints

\begin{equation}
\label{eq:d6truncationconstraints}
\hat{g}_{\underline{\sharp}\underline{\sharp}}=\alpha_{6}\, ,
\hspace{.5cm}
\hat{B}^{-}_{\mu\underline{\sharp}}=
-\sqrt{2} \hat{g}_{\mu\underline{\sharp}}\, ,
\hspace{.5cm}
\partial_{\underline{\sharp}}\hat{\epsilon}=0\, ,
\end{equation}

\noindent
can be consistently reduced in the direction $x^{\sharp}$ to a
solution of pure (Lorentzian $\alpha_{6}=-1$ or Euclidean
$\alpha_{6}=+1$) $N=1,d=5$ SUGRA whose bosonic action and gravitino
supersymmetry transformation rule are given by

\begin{equation}
\label{eq:N2d5SUGRAaction}
S={\displaystyle\int}
 d^{5}x\sqrt{|g|}\,\left[R
+{\textstyle\frac{1}{4}}\alpha_{6}F^{2}
-\alpha_{6}{\textstyle\frac{1}{12\sqrt{3}}}
{\textstyle\frac{\epsilon}{\sqrt{|g|}}}
FFV\right]\, . 
\end{equation}

\noindent
and

\begin{equation}
\label{eq:N=2d=5susyrule}
\delta_{\epsilon}\psi_{a}=
\left\{ \nabla_{a}
-{\textstyle\frac{i}{8\sqrt{3}}}\alpha_{6}^{3/2}
(\gamma^{bc}\gamma_{a}+2\gamma^{b}g^{c}{}_{a})
F_{bc}\right\}\epsilon\, .
\end{equation}

\noindent
with the same supersymmetries and with the same Killing spinor. The
5-dimensional fields are related to the 6-dimensional ones by

\begin{equation}
g_{\mu\nu}= \hat{g}_{\mu\nu}-\alpha_{6} 
\hat{g}_{\mu\underline{\sharp}} \hat{g}_{\nu\underline{\sharp}} \, ,
\hspace{1cm}
V_{\mu}= -\sqrt{3}\alpha_{6} \hat{g}_{\mu\underline{\sharp}}\, .
\end{equation}

Conversely, any solution of Lorentzian or Euclidean $N=1,d=5$ SUGRA
can be uplifted to a solution of Lorentzian $N=(1,0),\hat{d}=6$ SUGRA
whose fields are given by

\begin{equation}
\label{eq:5to6upliftingformulae}
  \begin{array}{rclrcl}
\hat{g}_{\underline{\sharp}\underline{\sharp}} & = & \alpha_{6}\, , &
\hat{B}_{\mu\underline{\sharp}} & = & 
-\sqrt{\textstyle\frac{2}{3}}\alpha_{6} V_{\mu}\, ,\\
& & & & & \\
\hat{g}_{\mu\underline{\sharp}} & = & 
\frac{1}{\sqrt{3}}\alpha_{6} V_{\mu}\, , 
\hspace{1cm}&
\hat{g}_{\mu\nu} & = & g_{\mu\nu}
+\frac{1}{3}\alpha_{6} V_{\mu}V_{\nu}\, .\\
\end{array}
\end{equation}


\section{Dimensional reduction of $N=1,\hat{d}=5$ SUGRA 
in the timelike direction}
\label{sec-dimred2}

The dimensional reduction of $N=1,\hat{d}=5$ SUGRA in one dimension
gives pure $N=2,d=4$ SUGRA (metric $e^{a}{}_{\mu}$, graviphoton
$V_{\mu}$ and a pair gravitinos that can be combined into a complex
Dirac spinor $\psi_{\mu}$) coupled to a vector multiplet (a vector
$W_{\mu}$, a pair of scalars $k$ and $l$ and a complex gaugino
$\lambda$) \cite{Chamseddine:1980sp}. As in the previous case, the
matter vector field $W_{\mu}$ is a combination of the scalars KK
vector field and of the 5-dimensional vector that has to be identified
studying the consistency of truncation in the action and supersymmetry
transformation rules.

We can reduce simultaneously both the Euclidean ($\alpha_{6}=+1$) and
Lorentzian ($\alpha_{6}=-1$) versions of $N=1,\hat{d}=5$ SUGRA and,
further, we can do it in timelike ($\alpha_{5}=+1$) and spacelike
($\alpha_{5}=-1$) directions in one shot\footnote{Only the case
  $\alpha_{6}=\alpha_{5}=+1$ is not possible, since we only have one
  time.}. The result will be a Euclidean theory when
$\alpha_{6}\alpha_{5}=-1$ and a Lorentzian theory when
$\alpha_{6}\alpha_{5}=+1$.  The reduction, identification of $W_{\mu}$
and consistent truncation of the vector multiplet are performed in
detail in Appendix~\ref{app-details} and we obtain the following
action and gravitino supersymmetry transformation rule

\begin{equation}
\label{eq:N2d4SUGRA}
S=\int d^{4}x \sqrt{|g|}\left[R
+{\textstyle\frac{1}{4}}\alpha_{6}F^{2}(V) \right]\, , 
\hspace{1cm}
\delta_{\epsilon}\psi_{a}=
\left\{ \nabla_{a} +{\textstyle\frac{i}{8}}\alpha_{6}^{3/2} 
\not\! F(V)\gamma_{a}\right\} \epsilon\, .
\end{equation}

These expressions exhibit a curious dependence on $\alpha_{6}$ that
indicates that we get two different Euclidean theories depending on
the order in which we have reduced in the spacelike and timelike
directions: if we compactify first in a spacelike direction
$\alpha_{6}=-1$ to Lorentzian $N=1,d=5$ SUGRA and then in a timelike
direction to Euclidean $N=2,d=4$ SUGRA we get the same theory that we
would obtain by Wick-rotating the Lorentzian theory considering the
vector field $V_{\mu}$ as a tensor.  If we compactify first in a
timelike direction $\alpha_{6}=+1$ to Euclidean $N=1,d=5$ SUGRA and
then in a spacelike direction to Lorentzian $N=2,d=4$ SUGRA we get the
same theory that we would obtain by Wick-rotating the Lorentzian
theory considering $V_{\mu}$ as a pseudotensor. These two Euclidean
theories that we denote by $N=2_{\alpha_{6}},d=4$ are then related by
an analytical continuation of the vector field $V_{\mu}\rightarrow i
V_{\mu}$. 

This apparent contradiction in the results of the two
compactifications in one timelike and one spacelike direction is
related to the well-known fact that Wick rotations and Hodge duality
(used in the reduction from $d=5$ to $d=4$) do not commute. 

The detailed calculations and results of the appendix imply that any
solution of Euclidean ($\alpha_{6}=+1$) or Lorentzian
($\alpha_{6}=-1$) $N=1,\hat{d}=5$ SUGRA satisfying the truncation
constraints

\begin{equation}
\hat{g}_{\underline{\sharp}\underline{\sharp}}=\alpha_{5}\, ,
\hspace{.5cm}
\hat{V}_{\underline{\sharp}}=0\, ,
\hspace{.5cm}
\sqrt{3}F(A)-{}^{\star}F(B) =0\, ,
\hspace{.5cm}
\partial_{\underline{\sharp}}\hat{\epsilon}=0\, ,
\end{equation}

\noindent
can be reduced, preserving all the unbroken supersymmetries, in the
direction $x^{\sharp}$ to a solution of pure Euclidean or Lorentzian
($\alpha_{5}\alpha_{6}=-1,+1$) $N=2_{\alpha_{6}},d=4$ SUGRA with
metric and vector field given by

\begin{equation}
g_{\mu\nu}=\hat{g}_{\mu\nu}-\alpha \hat{g}_{\mu\underline{\sharp}}  
\hat{g}_{\nu\underline{\sharp}} \, ,
\hspace{1cm}
V_{\mu}= -{\textstyle\frac{2}{\sqrt{3}}}\hat{V}_{\mu}\, .
\end{equation}

Conversely, any solution of pure Euclidean or Lorentzian
$N=2_{\alpha_{6}},d=4$ SUGRA can always be uplifted, preserving all
unbroken supersymmetries, to a solution of Euclidean ($\alpha_{6}=+1$)
or Lorentzian ($\alpha_{6}=-1$) $N=1,\hat{d}=5$ SUGRA whose fields are
related to the 4-dimensional ones by

\begin{equation}
\label{eq:4to5upliftingformulae}
  \begin{array}{rclrcl}
\hat{g}_{\underline{\sharp}\underline{\sharp}} & = & \alpha_{5}\, , &
\hat{V}_{\underline{\sharp}} & = & 0\, ,\\
& & & & & \\
\hat{g}_{\mu\underline{\sharp}} & = & \alpha_{5} A_{\mu}\, , 
\hspace{1cm}&
\hat{V}_{\mu} & = & -\frac{\sqrt{3}}{2} V_{\mu}\, ,\\
& & & & & \\
\hat{g}_{\mu\nu} & = & g_{\mu\nu}
+\alpha_{5} A_{\mu}A_{\nu}\, ,\hspace{1cm}& & & \\
\end{array}
\end{equation}

\noindent
where $A_{\mu}$ is defined in terms of $V_{\mu}$ by

\begin{equation}
F(A) = -{\textstyle\frac{1}{2}}{}^{\star}F(V)\, .  
\end{equation}

These are the essential formulae that we have to use for oxidation and
reduction of solutions and generalize those obtained in the Lorentzian
cases in \cite{Lozano-Tellechea:2002pn}.


\section{Maximally supersymmetric  vacua of Euclidean $N=1,d=5$ SUGRA}
\label{sec-vacuum5}

The most interesting solutions of $N=(1,0),\hat{d}=6$ SUGRA to which
the results of the previous sections may be applied are the maximally
supersymmetric vacua since their reduction should give \textit{all}
the maximally supersymmetric vacua of Euclidean and Lorentzian
$N=1,d=5$ SUGRA. The only maximally supersymmetric vacua of
$N=(1,0),\hat{d}=6$ SUGRA are
\cite{Gutowski:2003rg,Chamseddine:2003yy}

\begin{enumerate}
\item Minkowski spacetime.
  \item The 1-parameter family of Kowalski-Glikman (KG6) H$pp$-wave
    solutions found in Ref.~\cite{Meessen:2001vx}.
  \item The 1-parameter family of solutions with $aDS_{3} \times
    S^{3}$ geometry found in Ref.~\cite{Gibbons:1994vm} as the
    near-horizon limit of the self-dual string solution.
\end{enumerate}

From these three solutions, all the maximally supersymmetric vacua of
Lorentzian $N=1,d=5$ SUGRA, classified in \cite{Gauntlett:2002nw} can
be found by dimensional reduction in a spacelike direction
\cite{Lozano-Tellechea:2002pn,Gutowski:2003rg,Chamseddine:2003yy,Fiol:2003yq}
(see Figure~\ref{fig:d456vacua}). They are

\begin{enumerate}
\item Minkowski spacetime.
\item The 1-parameter family of Kowalski-Glikman $KG5$ H$pp$-wave
  solutions solutions found in Ref.~\cite{Meessen:2001vx}.
\item The 1-parameter family of solutions with $aDS_{3} \times S^{2}$
  geometry found in Ref.~\cite{Gibbons:1994vm} as near-horizon limit
  of the extreme string solution. It coincides with the near-horizon
  limit of the critically-rotating ($\jmath=1$) BMPV black hole
  \cite{Breckenridge:1996is}.
\item The 1-parameter family of solutions with $aDS_{2} \times S^{3}$
  geometry found in Ref.~ \cite{Kallosh:1996vy} as near-horizon limit
  of the extreme (non-rotating) black hole solution.
\item The 2-parameter family of $N=2, d=5$ solutions found in
  Ref.~\cite{Gauntlett:1998fz} as the near-horizon limit of the
  supersymmetric rotating BMPV black hole solution.
\item The 1-parameter family of G\"odel-like solutions found in
  \cite{Gauntlett:2002nw}.
\item A 2-parameter family of solutions found in
  \cite{Gauntlett:2002nw} that has just been identified as the
  over-rotating ($\jmath>1$) BMPV black holes \cite{Fiol:2003yq}.
\end{enumerate}

Now we want to find non-trivial maximally supersymmetric solutions of
Euclidean $N=1,d=5$ SUGRA by dimensional reduction in a spacelike
direction. As we are going to see, all of them can be obtained by Wick
rotation of maximally supersymmetric vacua of Lorentzian $N=1,d=5$
SUGRA.

Let us consider the spacelike dimensional reductions of 


\subsection{The $aDS_{3}\times S^{3}$ vacuum}

$aDS_{3}$ can be seen as a spacelike $U(1)$ fibration over $aDS_{2}$
with metric (for unit radius)

\begin{equation}
d\Pi_{(3)}^{2} = {\textstyle\frac{1}{4}}
\left[d\Pi_{(2)}^{2} -(d\eta +\textrm{sinh}\chi\, d\phi)^{2}  \right]\, ,
\end{equation}

\noindent
where

\begin{equation}
d\Pi_{(2)}^{2}= \textrm{cosh}^{2}\chi d\phi^{2}-d\chi^{2}\, ,  
\end{equation}

\noindent
is the metric of $aDS_{2}$ with unit radius, just as the round $S^{3}$
can be seen as a Hopf $U(1)$ fibration over $S^{2}$ with metric (for
unit radius)

\begin{equation}
d\Omega_{(3)}^{2} =  
{\textstyle\frac{1}{4}}\left[
d\Omega_{(2)}^{2}
+\left(d\psi +\cos{\theta}d\varphi\right)^{2}\right]\, ,
\end{equation}

\noindent
where 

\begin{equation}
d\Omega_{(2)}^{2} = d\theta^{2}+\sin^{2}{\theta} d\varphi^{2}\, ,  
\end{equation}

\noindent 
is the standard metric of $S^{2}$ with unit radius.

\begin{figure}[!ht]
\begin{center}
  \leavevmode \epsfxsize= 12cm 
\epsffile{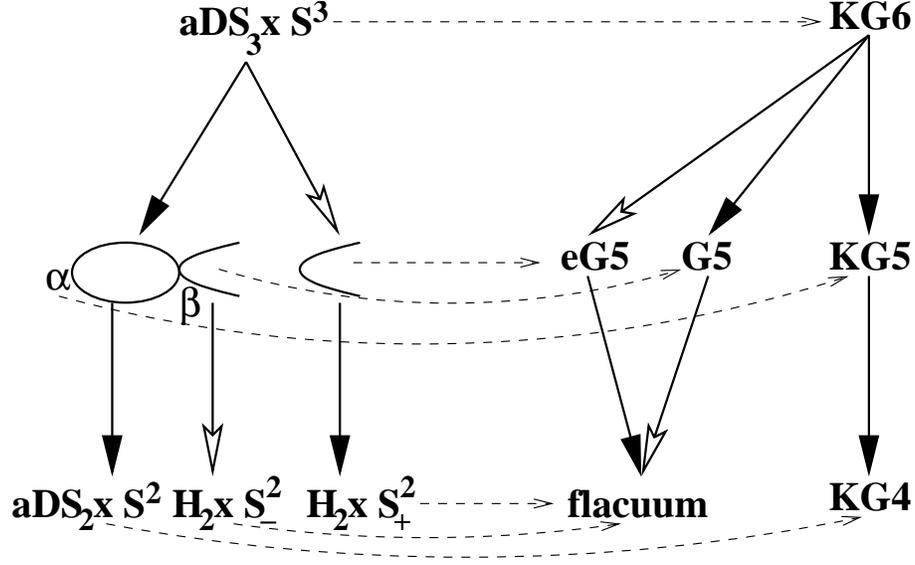}
\caption{Relations between the $d=4,5,6$ vacua with 8 supercharges. 
  Dimensional reduction in spacelike direction is represented by
  arrows with black arrowheads and in timelike directions with white
  arrowheads. The ellipse and hyperbolas represent the families of
  near-horizon limits of rotating and over-rotating extreme BMPV black
  holes. In particular, the elements $\alpha$ and $\beta$ of these
  families have the geometries of $aDS_{2}\times S^{3}$ and
  $aDS_{3}\times S^{2}$. Dotted lines indicate that the corresponding
  solutions are related by a Penrose-G\"uven-like limiting procedure.}
\label{fig:d456vacua}
\end{center}
\end{figure}

It is possible to perform spacelike dimensional reductions on linear
combinations of these two fibers (i.e.~of the coordinates $\eta$ and
$\psi$), obtaining the near-horizon limit of the extreme rotating BMPV
black hole \cite{Lozano-Tellechea:2002pn} (see
Figure~\ref{fig:d456vacua}). The metrics of all these 5-dimensional
solutions are $U(1)$ fibrations over $aDS_{2}\times S^{2}$, the $U(1)$
fiber being orthogonal to the one we have reduced over.

It is also possible to see $aDS_{3}$ as a timelike $U(1)$ fibration
over the hyperbolic plane $H_{2}$ with metric (for unit radius)

\begin{equation}
\label{eq:timelikefibration}
d\Pi_{(3)}^{2} = {\textstyle\frac{1}{4}}
\left[(dt +\textrm{cosh}\chi\, d\phi)^{2} -dH_{(2)}^{2}  \right]\, ,
\end{equation}

\noindent
where

\begin{equation}
dH_{2}^{(2)} = {\rm sinh}^{2}\chi d\phi^{2} +d\chi^{2}\, ,
\end{equation}

\noindent
is the standard metric of the hyperbolic plane of unit radius.  We can
perform further spacelike and also timelike dimensional reductions on
linear combinations of these two fibers. The spacelike reductions give
the near-horizon limit of the extreme over-rotating BMPV black hole
\cite{Fiol:2003yq} and here we are going to see that the timelike
reductions give their analytical continuations to Euclidean signature.

The maximally supersymmetric solution with $aDS_{3}\times S^{3}$
geometry is given by

\begin{equation}
\left\{
  \begin{array}{rcl}
d\hat{s}^{2} & = &  R_{3}^{2}d\Pi_{(3)}^{2} - R_{3}^{2}d\Omega_{(3)}^{2}\, ,\\
& & \\
\hat{H}^{-} & = & 4\sqrt{2}R_{3}\, [\omega_{aDS_{3}}-\omega_{(3)}]\, ,\\
  \end{array}
\right.
\end{equation}

\noindent
where $\omega_{aDS_{3}}$ and $\omega_{(3)}$ are the volume forms of
the unit radius $aDS_{3}$ and $S^{3}$ metrics written above. Using the
timelike $U(1)$ fibration Eq.~(\ref{eq:timelikefibration}) we find
that a convenient gauge for the 2-form potential is 

\begin{equation}
\hat{B}^{-} = 
-R_{3}/\sqrt{2} \left[\textrm{cosh}\chi d\phi\wedge dt 
-\cos\theta d\varphi \wedge d\psi \right]\, .  
\end{equation}

\noindent
Rescaling $t$ and $\psi$ by $2/R_{3}$ and boosting the metric in
the direction $\psi$

\begin{equation}
  \begin{array}{rcl}
t & = & 2/R_{3}(\textrm{cosh}\xi\, t^{\prime} 
+\textrm{sinh}\xi\, \psi^{\prime})\, ,\\
& & \\
\psi & = & 2/R_{3}(\textrm{sinh}\xi\, t^{\prime} 
+\textrm{cosh}\xi\, \psi^{\prime})\, ,\\
  \end{array}
\end{equation}

\noindent
the solution takes the form

\begin{equation}
\left\{
  \begin{array}{rcl}
d\hat{s}^{2} & = &  
(dt^{\prime} +A^{(t)})^{2} -(d\psi^{\prime} +A^{(\psi)})^{2} 
-(R_{3}/2)^{2}[dH_{(2)}^{2} +d\Omega_{(2)}^{2}]\, ,\\
& & \\
\hat{B}^{-} & = & -\sqrt{2}A^{(t)}\wedge dt^{\prime} 
+\sqrt{2}A^{(\psi)}\wedge d\psi^{\prime}\, , \\
  \end{array}
\right.
\end{equation}

\noindent
where $A^{(t,\psi)}$ are the 1-forms

\begin{equation}
  \begin{array}{rcl}
A^{(t)} & =  & R_{3}/2(\textrm{cosh}\xi \textrm{cosh}\chi d\phi
-\textrm{sinh}\xi\cos\theta d\varphi)\, ,\\
& & \\
A^{(\psi)} & = & R_{3}/2(\textrm{cosh}\xi\cos\theta d\varphi
-\textrm{sinh}\xi\textrm{cosh}\chi d\phi)\, .\\
\end{array}
\end{equation}

The truncation constraints Eqs.~(\ref{eq:d6truncationconstraints}) are
satisfied for $x^{\sharp}=t^{\prime}$ and $x^{\sharp}=\psi^{\prime}$
and, thus, we can dimensionally reduce this solution to a maximally
supersymmetric solution of Euclidean or Lorentzian $N=1,d=5$ SUGRA.
In the spacelike case we get 

\begin{equation}
\label{eq:overrotatingBMPV}
\left\{
  \begin{array}{rcl}
ds^{2} & = &  
(dt^{\prime} +A^{(t)})^{2} 
-(R_{3}/2)^{2}[dH_{(2)}^{2} +d\Omega_{(2)}^{2}]\, ,\\
& & \\
V & = & -\sqrt{3} A^{(\psi)}\, , 
\hspace{1cm}
F = -\frac{\sqrt{3}}{2}R_{3}(\textrm{sinh}\xi \omega_{aDS_{2}} 
+\textrm{cosh}\xi \omega_{(2)})\, ,\\
  \end{array}
\right.
\end{equation}

\noindent
which is the near-horizon limit of the extreme critically rotating and
over-rotating BMPV black hole \cite{Fiol:2003yq}.  $\jmath={\rm
  cosh}\xi$ is the rotation parameter. The critically rotating case
has $\jmath=1$, i.e.~${\rm sinh}\xi=0$ and the fibration takes place
only over the hyperbolic plane, giving $aDS_{3}$. The total metric
describes the solution $aDS_{3}\times S_{2}$. For any other value of
$\xi$, $\jmath >1$, and we are in the over-rotating case.

In the timelike case, we get

\begin{equation}
\label{eq:EuclideanrotatingBMPV}
\left\{
  \begin{array}{rcl}
ds^{2} & = &  
-(d\psi^{\prime} +A^{(\psi)})^{2} 
-(R_{3}/2)^{2}[dH_{(2)}^{2} +d\Omega_{(2)}^{2}]\, ,\\
& & \\
V & = & -\sqrt{3} A^{(t)}\, ,
\hspace{1cm}
F = -\frac{\sqrt{3}}{2}R_{3}(\textrm{cosh}\xi \omega_{H_{2}} 
+\textrm{sinh}\xi \omega_{(2)})\, .\\
  \end{array}
\right.
\end{equation}

These two solutions are related by a Wick rotation $t\rightarrow
i\psi$ $V\rightarrow iV$ which has to be accompanied of
$\xi\rightarrow \xi -i\pi/2$ to make $V$ real again.  The Euclidean
solution can also be obtained by a Wick rotation $t\rightarrow i\psi$
$V\rightarrow iV$ from the near-horizon limit of the extreme BMPV
black hole accompanied of $\xi\rightarrow i\xi$ to make $V$ real
again.

On the other hand, these solutions can be seen, respectively, as a
timelike $\mathbb{R}$ or spacelike $U(1)$ fibration over $H_{2}\times
S^{2}$.

Let us now consider the timelike dimensional reductions of


\subsection{The $KG6$ H$pp$-wave vacuum}

As shown in \cite{Lozano-Tellechea:2002pn}, the KG6 solution can be
conveniently written substituting $g_{uu}$ completely by a Sagnac
connection 1-form $\omega=\lambda(x^{1}dx^{2}-x^{3}dx^{4})$ whose main
property is that $d\omega$ is anti-selfdual in the 4-dimensional
Euclidean space spanned by $\vec{x}_{4}=(x^{1},x^{2},x^{3},x^{4})$:

\begin{equation}
\left\{
  \begin{array}{rcl}
d\hat{s}^{2} & = & 2du(dv + \sqrt{2}\omega) -d\vec{x}_{4}^{2}\, ,\\
& & \\
\hat{B}^{-} & = & -2\omega\wedge du\, ,\\
  \end{array}
\right.
\end{equation}

\noindent
and, using $u=(t+z)/sqrt{2}$, we can rewrite it in the form

\begin{equation}
\left\{
  \begin{array}{rcl}
d\hat{s}^{2} & = & (dt+\omega)^{2}-(dz-\omega)^{2} -d\vec{x}_{4}^{2}\, ,\\
& & \\
\hat{B}^{-} & = & -\sqrt{2}\omega\wedge dt-\sqrt{2}\omega\wedge dz  \, ,\\
  \end{array}
\right.
\end{equation}

\noindent
that shows that the truncation constraints
Eqs.~(\ref{eq:d6truncationconstraints}) are satisfied for
$x^{\sharp}=t$ and $x^{\sharp}=z$ and, thus, we can dimensionally
reduce this solution to maximally supersymmetric solutions of
Euclidean or Lorentzian $N=1,d=5$ SUGRA. The spacelike dimensional
reduction gives the G\"odel solution

\begin{equation}
\label{eq:Goedel}
\left\{
  \begin{array}{rcl}
ds^{2} & = & (dt+\omega)^{2} -d\vec{x}_{4}^{2}\, ,\\
& & \\
V & = & -\sqrt{3}\omega \, ,\\
  \end{array}
\right.
\end{equation}

\noindent
and the timelike reduction gives its Euclidean version

\begin{equation}
\label{eq:EuclideanGoedel}
\left\{
  \begin{array}{rcl}
ds^{2} & = & -(dz-\omega)^{2} -d\vec{x}_{4}^{2}\, ,\\
& & \\
V & = & -\sqrt{3}\omega \, ,\\
  \end{array}
\right.
\end{equation}

\noindent
that can be obtained by Wick-rotating $t\rightarrow iz$, $V\rightarrow
iV$ plus $\lambda\rightarrow -i\lambda$ to make $V$ real.  These two
solutions can also be seen as timelike $\mathbb{R}$ or spacelike
$U(1)$ fibrations over the 4-dimensional flat Euclidean space.

No more timelike dimensional reductions satisfying the truncation
constraints seem to be possible, and, thus, we do not expect any more
maximally supersymmetric solutions of Euclidean $N=1,d=5$ SUGRA.


\section{Maximally supersymmetric  vacua of Euclidean $N=2_{\pm},d=4$ SUGRA}
\label{sec-vacuum4}

Next, we are going to dimensionally reduce the maximally
supersymmetric vacua of Euclidean and Lorentzian $N=1,d=5$ SUGRA to
find maximally supersymmetric vacua of Euclidean $N=2_{\pm},d=4$
SUGRAs. The maximally supersymmetric vacua of Lorentzian $N=2,d=4$
SUGRA\footnote{All the solutions of Lorentzian $N=2,d=4$ SUGRA that
  preserve some supersymmetry were found in \cite{Tod:pm}. The
  maximally supersymmetric vacua of the theory are just three:
  Minkowski spacetime, the dyonic Robinson-Bertotti solution
  \cite{kn:RB} and the KG4 H$pp$-wave solution
  \cite{Kowalski-Glikman:im}.} have been obtained by spacelike
dimensional reduction of the vacua of Lorentzian $N=2,d=5$ SUGRA
listed in Section~\ref{sec-vacuum5} \cite{Lozano-Tellechea:2002pn}.
 
The only Lorentzian solutions that satisfy the truncation constraints
for timelike reduction seem to be Minkowski spacetime, the
near-horizon (NH) limit of the critically rotating and over-rotating
BMPV black holes (BH) Eq.~(\ref{eq:overrotatingBMPV}) and the G\"odel
solution Eq.~(\ref{eq:Goedel}).  The only Euclidean solutions that
satisfy the analogous truncation constraints are the Euclidean NH
limit of the rotating BMPV BH\footnote{As we have seen, the
  Wick-rotated NH limits of the extreme over-rotating and un
  ``under-rotating'' BMPV BHs are identical.}
Eq.~(\ref{eq:EuclideanrotatingBMPV}) and the Euclidean G\"odel
solution Eq.~(\ref{eq:EuclideanGoedel}) Let us study the non-trivial
cases.


\subsection{NH limits of the extreme critically rotating and 
  over-rotating BMPV BHs}

Using the formulae of Section~\ref{sec-dimred2} on the solution
Eq.~(\ref{eq:overrotatingBMPV}) we immediately get a maximally
supersymmetric solution of Euclidean $N=2_{-},d=4$ SUGRA

\begin{equation}
\label{eq:H2S2-}
\left\{
  \begin{array}{rcl}
ds^{2} & = &  -(R_{3}/2)^{2}[dH_{2}^{2}+d\Omega_{2}^{2}]\, ,\\
& & \\
F & = & R_{3}\left({\rm sinh}\xi \omega_{H_{2}} 
+{\rm cosh}\xi \omega_{(2)}\right)\, ,\\
  \end{array}
\right.  
\end{equation}

\noindent 
which is the Wick-rotated ($\phi=-it$) version of the well-known
dyonic Robinson-Bertotti solution \cite{kn:RB}. Observe that, in order
to have a real Euclidean solution, the Lorentzian electric-magnetic
rotation angle has also been Wick-rotated $\xi\rightarrow i\xi$.


\subsection{The Euclidean NH limits of extreme rotating BMPV BHs}

Using the formulae of Section~\ref{sec-dimred2} on the solution
Eq.~(\ref{eq:EuclideanrotatingBMPV}) we immediately get a maximally
supersymmetric solution of Euclidean $N=2_{+},d=4$ SUGRA

\begin{equation}
\label{eq:H2S2+}
\left\{
  \begin{array}{rcl}
ds^{2} & = &  -(R_{3}/2)^{2}[dH_{2}^{2}+d\Omega_{2}^{2}]\, ,\\
& & \\
F & = & R_{3}\left({\rm cosh}\xi \omega_{H_{2}} 
+{\rm sinh}\xi \omega_{(2)}\right)\, ,\\
  \end{array}
\right.  
\end{equation}

This solution is related to the Euclidean dyonic Robinson-Bertotti
solution by analytical continuation of the vector field $V\rightarrow
iV$ together with $\xi\rightarrow \xi -i\pi/2$.


\subsection{The Euclidean and Lorentzian G\"odel solutions and the 
flacuum solution}

The $\hat{d}=5$ G\"odel solutions are given in Eqs.~(\ref{eq:Goedel})
and (\ref{eq:EuclideanGoedel}).  The formulae of
Section~\ref{sec-dimred2} give, in both cases, the remarkable
maximally supersymmetric 4-dimensional vacuum solution with constant
anti-selfdual flux (\textit{flacuum}) found in \cite{Berkovits:2003kj}

\begin{equation}
\label{eq:flacuum}
\left\{
  \begin{array}{rcl}
ds^{2} & = &  -d\vec{x}^{\, 2}_{4}\, ,\\
& & \\
V & = & 2\omega\, .\\
  \end{array}
\right.  
\end{equation}

The triviality of the metric in presence of non-trivial matter fields
is surprising at first sight, but it is not the first example of
non-trivial Euclidean SUGRA solution with flat spacetime: the
Einstein-frame D-instanton metric \cite{Gibbons:1995vg} (and that of
its 4-dimensional analogue \cite{Giddings:1989bq}) is also flat. In
all these cases, the Euclidean version of the energy-momentum tensor
vanishes identically, but for different reasons. In the D-instanton
case the dilaton and RR 0-form energy-momentum tensors cancel each
other on-shell. In the present case, anti-selfduality of the vector
field strength makes the energy-momentum tensor vanish identically.
This can be easily seen using
Eq.~(\ref{eq:vectorenergymomentumtensor}) which tells us that,
actually, the Euclidean version of energy-momentum tensor for any
(anti-) selfdual Abelian or non-Abelian vector field vanishes
identically and, thus, decouples from the metric\footnote{But not the
  other way around since (anti-) selfduality is established with
  respect to a given metric.}.  This implies that solutions such as
the BPST instanton \cite{Belavin:fg} are not just solutions of the
Euclidean Yang-Mills equations on $S^{4}$ with the standard metric,
but they are also solutions of the full supergravity equations,
provided that the metric satisfies the vacuum Einstein equations. We
will discuss this and similar solutions in the next section.

The above solution is guaranteed to be maximally supersymmetric in
both Euclidean $N=2_{\pm},d=4$ SUGRA. It is worth studying how the
Killing spinors integrability conditions are satisfied. They take the
form\footnote{For simplicity we consider the $\alpha_{6}=-1$ case.}

\begin{equation}
\left\{-{\textstyle\frac{1}{4}}R_{\mu\nu}{}^{ab}\gamma_{ab}  
+{\textstyle\frac{1}{4}}T_{[\mu|\rho|}\gamma^{\rho}{}_{\nu]}  
+{\textstyle\frac{1}{4}}\not\!\nabla 
\left( F_{\mu\nu} + {}^{\star}F_{\mu\nu} \gamma^{\sharp}\right)
\right\} \epsilon =0\, .
\end{equation}

In the case at hand the first term vanishes identically because the
space is flat, the second because of the anti-selfduality of the
vector field strength and the third and fourth because the field
strength is covariantly constant (constant in Cartesian coordinates).

(Anti-) selfduality implies flatness and, then, covariant constancy
determines a unique solution (up to $SO(4)$ rotations). Suppressing
this last requirement allows for more solutions\footnote{For instance,
  $A=dx^{\sharp}/r +\cos{\theta}d\varphi$ in spherical
  coordinates.} which would only preserve a half of the
supersymmetries.




This solution preserves all supersymmetries and the Killing spinors
depend on all coordinates:

\begin{equation}
\epsilon (x) ={\rm exp} \{-{\textstyle\frac{1}{8}}\not\! F\!\!\not\! x \}
\epsilon_{0}\, ,  
\end{equation}

\noindent
where $\epsilon_{0}$ is a constant spinor and $\not\! x \equiv
x^{\mu}\gamma_{\mu}$. Using the anti-selfduality of $F$, we can
rewrite the exponent in this form

\begin{equation}
-{\textstyle\frac{1}{8}}\not\! F\!\!\not\! x =  -{\textstyle\frac{1}{4}}
\not\! F_{\mu}x^{\mu} (1-\gamma^{\sharp})\, ,
\end{equation}

\noindent
where $(\gamma^{\sharp})^{2}=1$ in this case. This implies that
$(\not\! F\!\!\not\! x)^{2}=0$ and, thus, the Killing spinors can be
written in the form

\begin{equation}
\epsilon (x) = \{1 -{\textstyle\frac{1}{4}}
\not\! F_{\mu}x^{\mu} (1-\gamma^{\sharp}) \}
\epsilon_{0}\, .
\end{equation}

If we split our complex Dirac spinors into chiral halves, we see that
the positive chirality Killing spinor is simply constant, as in empty
Euclidean space, and the negative chirality Killing spinor has a
$x$-dependent deformation with respect to that of empty Euclidean
space which induces a deformation of the supersymmetry algebra
\cite{Seiberg:2003yz,Berkovits:2003kj} that affects only to
negative-chirality objects. Using the methods of
\cite{Figueroa-O'Farrill:1999va} and the notation of
\cite{Ortin:2002qb,Alonso-Alberca:2002gh}, and using 4-component
complex (Dirac) spinors, we find the following non-vanishing (anti-)
commutators

\begin{equation}
  \begin{array}{rcl}
\left\{\mathcal{Q}^{\dagger}_{(\alpha)},\mathcal{Q}_{(\beta)}\right\} & = & 
(\gamma^{1}\gamma^{a})_{\alpha\beta} P_{(a)} 
-[\gamma^{1} {\textstyle\frac{1}{2}}
(1-\gamma^{\sharp})]_{\alpha\beta}M\, , \\
& & \\
\left[ \mathcal{Q}_{(\alpha)}, P_{(a)} \right] & = & 
 -\mathcal{Q}_{(\beta)}\Gamma_{s}(P_{(a)})^{\beta}{}_{\alpha} \, ,\\
& & \\
\left[ {\cal Q}_{(\alpha)},M\right] & = & 
-\mathcal{Q}_{(\beta)} \Gamma_{s}(M)^{\beta}{}_{\alpha}\, ,\\
& & \\
\left[ P_{(a)},M \right] & = & -P_{(b)} \Gamma_{v}(M)^{b}{}_{a}\, .\\
\end{array}
\end{equation}
 
The bosonic generators $P_{(a)}$ of this algebra are associated to the
translational Killing vectors $\partial_{a}$. The bosonic generator
$M$ is a Lorentz ($SO(4)$) rotation $\frac{1}{2}F^{ab}M_{ab}$ where
the Lorentz generators $M_{ab}$ are associated the Killing vectors
$-2x_{[a}\partial_{b]}$. Both the $P_{a}$s and $M$ act on spinors
through their spinorial representations

\begin{equation}
\label{eq:spinorialgenerators}
\Gamma_{s}(P_{(a)})^{\beta}{}_{\alpha} = \left[{\textstyle\frac{1}{4}}
\not\! F_{a}(1-\gamma^{\sharp})\right]^{\beta}{}_{\alpha}\, ,
\hspace{1cm}
\Gamma_{s}(M)^{\beta}{}_{\alpha}= \left[{\textstyle\frac{1}{2}}\not\! F
\right]^{\beta}{}_{\alpha}=
\left[{\textstyle\frac{1}{4}}\not\! F (1+\gamma^{\sharp}
\right]^{\beta}{}_{\alpha}\, ,
\end{equation}

\noindent
and on vectors through the vector representation

\begin{equation}
\Gamma_{v}(M)^{b}{}_{a}=-F^{b}{}_{a}\, .  
\end{equation}

This superalgebra is a deformation of the standard Poincar\'e
$N=2,d=4$ superalgebra. This is a common property of all the maximally
supersymmetric vacua of $N=2,d=4$ SUGRA.  However, this deformation is
particularly interesting because the bosonic translations algebra is
not modified and, actually, only the anticommutator of the negative
chirality part of the supercharge is deformed and this deformation
induces a deformation of the anticommutator of the fermionic
superspace coordinates of negative chirality
\cite{Seiberg:2003yz,Berkovits:2003kj}.

The relation between the flacuum and G\"odel solutions (whose Killing
spinors are, according to our results, absolutely identical) implies a
relation between their superalgebras which is important to understand.
The only difference between them is the occurrence in the G\"odel
superalgebra of the bosonic generator $P_{(0)}$ associated to time
translations in the anticommutator of the supercharges and, more
importantly, in the commutator of the generators associated to space
translations $P_{(a)}$. If we write the G\"odel solution
Eq.~(\ref{eq:Goedel}) in the gauge in which the 1-form $\omega$ takes
the form

\begin{equation}
\omega= {\textstyle\frac{1}{2}}F_{ab}x^{a}dx^{b}\, ,\,\,\,\,
a,b=1,2,3,4\, ,\,\,\,\,
F_{12}=-F_{34}=\lambda\, ,
\end{equation}

\noindent
we find the Killing vectors\footnote{There are more Killing vectors,
  but they are not relevant insofar they do not appear in the
  anticommutator of the supercharges.}

\begin{equation}
\partial_{t}\, ,
\hspace{1cm}
\partial_{a} -{\textstyle\frac{1}{2}}F_{ab}x^{b}\partial_{t}\, ,  
\hspace{1cm}
-{\textstyle\frac{1}{2}}F^{ab}x_{a}\partial_{b}\, ,
\end{equation}

\noindent
associated, respectively, to the generators of time and space
translations $P_{(0)}$ and $P_{(a)}$ and to the rotation $M$. The
non-vanishing (anti-) commutators of the G\"odel superalgebra are

\begin{equation}
  \begin{array}{rcl}
\left\{\mathcal{Q}^{\dagger}_{(\alpha)},\mathcal{Q}_{(\beta)}\right\} & = & 
(\gamma^{1}\gamma^{a})_{\alpha\beta} P_{(a)} 
+(\gamma^{1}\gamma^{\sharp})_{\alpha\beta} P_{(0)} 
-[\gamma^{1} {\textstyle\frac{1}{2}}
(1-\gamma^{\sharp})]_{\alpha\beta}M\, , \\
& & \\
\left[ \mathcal{Q}_{(\alpha)}, P_{(a)} \right] & = & 
 -\mathcal{Q}_{(\beta)}\Gamma_{s}(P_{(a)})^{\beta}{}_{\alpha} \, ,\\
& & \\
\left[ {\cal Q}_{(\alpha)},M\right] & = & 
-\mathcal{Q}_{(\beta)} \Gamma_{s}(M)^{\beta}{}_{\alpha}\, ,\\
& & \\
\left[ P_{(a)},M \right] & = & -P_{(b)} \Gamma_{v}(M)^{b}{}_{a}\, ,
\hspace{1cm}
\left[P_{(a)},P_{(b)}\right]=F_{ab}P_{(0)}\, .\\
\end{array}
\end{equation}

In this background space translations commute only up to a time
translation, because space translations $\delta x^{a}$ have to be
compensated by a time translation $\delta t= \frac{1}{2}F^{ab}x_{b}$
to leave the metric invariant. This deformation is interesting because
is suggests that string theory quantized in the G\"odel background
will give a non-commutative theory in which, instead of modifying the
commutator of positions, it is the commutator of momenta which are
deformed as above. Supersymmetry requires a corresponding modification
of the anticommutator f the supercharges as in the flacuum solution.

In the dimensional reduction to the flacuum solution the generator
$P_{(0)}$ becomes a central charge that generates gauge
transformations of the Kaluza-Klein vector field. Actually, it is easy
to check that space translations only leave the vector field of the
flacuum solution up to gauge transformations with gauge parameter
$\frac{1}{2}F^{ab}x_{b}$. In this sense, the flacuum superalgebra is
not complete and one should add $P_{(0)}$s as in the G\"odel
superalgebra. However, if we are considering only $t$-independent
5-dimensional configurations which correspond to solutions with zero
$P_{(0)}$ charge, the difference is immaterial. In $N=2,d=4$ SUGRA
there are no charged perturbative states in $N=2,d=4$ SUGRA, but
non-perturbative states that would be associated to 5-dimensional KK
modes with non-trivial time dependence are charged with respect to
$P_{(0)}$ and would feel the non-commutativity of the momenta.


\subsection{The flacuum solution as an Abelian instanton}

Another interesting aspect of the flacuum solution is its possible
interpretation as an instanton. This is only possible if the space is
compactified into a $T^{4}$ to make the action integral finite. This
is, also, a base space on which the $U(1)$ gauge group can be
non-trivially fibered. Euclidean Yang-Mills solutions on $T^{4}$ have
been thoroughly studied in the literature\footnote{See, for instance
  the pedagogical review \cite{Gonzalez-Arroyo:1997uj} whose notation
  we will loosely follow. In particular we use the period vectors
  $(\hat{a})^{c}\equiv \delta_{(a)}{}^{c}l^{(a)}$, where
  summation is not to be made over indices within parenthesis.} and we
are simply going to apply the known results to the present case.

To compactify the solution on $T^{4}$ we take the quotient of
$\mathbb{R}^{4}$ by the $\mathbb{Z}^{4}$ Abelian group of discrete
translations along the four coordinates $x^{a}$ with periods
$l^{a}$. The vector field of our solution, which we rewrite for
convenience in a new gauge

\begin{equation}
V= \lambda(x^{1}dx^{2}-x^{2}dx^{1}-x^{3}dx^{4}+x^{4}dx^{3})\equiv 
F_{ab}x^{a}dx^{b}\, ,  
\end{equation}

\noindent
is not strictly periodic on $T^{4}$: when we move around the $a$-th
period from $x$ to $x+\hat{a}$ it changes by a gauge transformation

\begin{equation}
V(x+\hat{a}) = V(x) +d\Lambda_{a}(x)\, ,
\hspace{1cm}
\Lambda_{a}(x)=l^{(a)}F_{(a)b}x^{b}\, ,
\end{equation}

\noindent
where $\Lambda_{a}(x)$ are the $U(1)$ parameters, defined modulo
$2\pi$.  Consistency requires that
$V(x+\hat{a}+\hat{b})=V(x+\hat{b}+\hat{a})$, that is

\begin{equation}
\Lambda_{a}(x+\hat{b})+\Lambda_{b}(x)=
\Lambda_{b}(x+\hat{a})+\Lambda_{a}(x)\,\,\,\, {\rm mod}(2\pi)\, ,
\end{equation}

\noindent
which in our case implies

\begin{equation}
\lambda l^{1}l^{2} =\pi n\, ,
\hspace{1cm}
\lambda l^{3}l^{4} =\pi m\, ,
\end{equation}

\noindent
for two integers $n,m$ that label the possible non-trivial bundles,
the topological number being given by their product $nm$. The
Euclidean action of these SUGRA solutions is, therefore, in our
conventions

\begin{equation}
S= -4\pi^{2}|nm|\, .
\end{equation}

Let us consider now taking the quotient on the spinor bundle, which
requires making the identifications

\begin{equation}
\epsilon(x+\hat{a}) = 
{\rm exp} 
\{-{\textstyle\frac{l^{(a)}}{8}}\not\! F\! 
\gamma_{(a)} \}\epsilon(x)\sim \epsilon(x)\, ,
\end{equation}

\noindent
consistently. These transformations are not $SO(4)$ rotations but are
still elements of the bigger holonomy group of the supercovariant
derivative of $N=2,d=4$ SUGRA, which, by arguments similar to those in
\cite{Hull:2003mf,Papadopoulos:2003jk} can be shown to be
$SL(4,\mathbb{R})$, and, in this sense, from this more general point
of view, they should be admissible.

On the other hand, the consistency of these identifications is ensured
by the fact that the above transformations are indeed a representation
of the commutative $\mathbb{Z}_{4}$ group of discrete translations:
the generators of translations in the $x^{a}$ directions, $P_{(a)}$
are represented on the spinors by the mutually commuting operators
$\Gamma_{s}(P_{(a)})$ given in Eq.~(\ref{eq:spinorialgenerators}).

The consistency of this construction has interesting implications: the
compactness of the gauge group ($U(1)$ instead of $\mathbb{R}$),
reflected by the periodicity of the gauge parameters
$\Lambda_{a}(x)$ implies the compactness of the $\hat{d}=5$ time
coordinate. All the vacua of $N=1,\hat{d}=5$ SUGRA (except the KG5
H$pp$-wave) seem to be non-trivial spacelike or timelike fibrations
over 4-dimensional Euclidean space or over $H_{2}\times S^{2}$.

In this respect, it is remarkable that the only solutions that can be
reduced consistently in the timelike direction (preserving all
supersymmetries) have closed timelike curves or (in the $\jmath=1$
closed lightlike curves.


\section{(Anti-) selfdual  vacua in Euclidean SUGRA}
\label{sec-selfdual}

It is not difficult to construct generalizations of the flacuum
solution.  Let us consider a $(p+1)$ form potential $A_{(p+1)}$ with
field strength $F_{(p+2)}=d A_{(p+1)}$ coupled to $d$-dimensional
gravity:

\begin{equation}
S= \int d^{d}x\sqrt{|g|}\left[R 
-{\textstyle\frac{(-1)^{p}}{2\cdot (p+2)!}}F^{2}_{(p+2)}\right]\, .  
\end{equation}

The Einstein equation is

\begin{equation}
R_{\mu\nu}-{\textstyle\frac{1}{2}}g_{\mu\nu}R = 
{\textstyle\frac{(-1)^{p}}{2\cdot (p+1)!}} T^{(p+1)}_{\mu\nu}\, ,
\end{equation}

\noindent
where the energy-momentum tensor is given by 

\begin{equation}
\label{eq:energymomentum1}
T^{(p+1)}_{\mu\nu}= F_{(p+2)\, \mu}{}^{\rho_{1}\cdots\rho_{p+1}}  
F_{(p+2)\, \nu\rho_{1}\cdots\rho_{p+1}} 
-{\textstyle\frac{1}{2\cdot(p+2)}}g_{\mu\nu} F^{2}_{(p+2)}\, ,
\end{equation}

\noindent
or, equivalently, by the more useful expression

\begin{equation}
\label{eq:energymomentum2}
T^{(p+1)}_{\mu\nu}= {\textstyle\frac{1}{2}} 
\left[ F_{(p+2)\, \mu}{}^{\rho_{1}\cdots\rho_{p+1}}  
F_{(p+2)\, \nu\rho_{1}\cdots\rho_{p+1}}
-\alpha {}^{\star}F_{(p+2)\, \mu}{}^{\rho_{1}\cdots\rho_{\tilde{p}+1}}  
{}^{\star}F_{(p+2)\, \nu\rho_{1}\cdots\rho_{\tilde{p}+1}} \right]\, ,
\end{equation}

\noindent
where $\tilde{p}=d-p-4$.

We are interested in solutions with vanishing energy-momentum tensor.
A necessary condition for the energy-momentum tensor to vanish is that
its trace does. Using (\ref{eq:energymomentum1}) we find that 

\begin{equation}
T^{(p+1)}_{\mu}{}^{\mu}\sim 
\left[1-{\textstyle\frac{d}{2(p+2)}}\right] F^{2}_{(p+2)}\, ,
\end{equation}

\noindent
which vanishes when $F^{2}_{(p+2)}=0$ or when $p+2=d/2$. For
non-trivial potentials, the first case is only possible in Lorentzian
signature, but we know that the Lorentzian energy-momentum tensor only
vanishes for vanishing field strengths. Thus, a vanishing
energy-momentum tensor is only possible in Euclidean signature for
$d=2(p+2)$ and, using (\ref{eq:energymomentum2}), it requires

\begin{equation}
F_{(p+2)\, \mu}{}^{\rho_{1}\cdots\rho_{p+1}}  
F_{(p+2)\, \nu\rho_{1}\cdots\rho_{p+1}}
={}^{\star}F_{(p+2)\, \mu}{}^{\rho_{1}\cdots\rho_{p+1}}  
{}^{\star}F_{(p+2)\, \nu\rho_{1}\cdots\rho_{p+1}}\, ,
\end{equation}

\noindent
which can be solved in any even dimension: in $d=4n$ using (anti-)
selfdual field strengths (as we have seen in $d=4$) and in $d=4n+2$
using configurations which would be (anti-) self-dual in Lorentzian
signature. For example, in $d=6$ we could use $F_{(3)\, 123}= F_{(3)\,
  456}= \lambda$ which has $F_{(3)\, 123}= -{}^{\star}F_{(3)\, 123}$
but $F_{(3)\, 123}= +{}^{\star}F_{(3)\, 123}$ (i.e.~not definite
selfduality properties) but clearly satisfies the above condition.

Unfortunately, these simple \textit{Ans\"atze} for flacuum solutions
do not work in presence of scalars such as the string dilaton that
couple to $F^{2}_{(p+2)}$. Still, since the sign of the Euclidean
``kinetic'' terms and energy-momentum tensors is not uniquely
defined\footnote{They depend on whether the Lorentzian field is a
  tensor or a pseudotensor, as the KK vector generically turns out to
  be.}, it is possible that the energy-momentum tensors and their
contributions to the source for the dilaton of two fields cancel each
other.  Examples of these solutions are provided by the dimensional
reduction in the time direction of the supersymmetric G\"odel
solutions of $N=1,d=11$ SUGRA found in \cite{Harmark:2003ud}. These
solutions take the form

\begin{equation}
\left\{
  \begin{array}{rcl}
d\hat{s}^{2} & = & (dt +c_{i}\omega^{i})^{2} -d\vec{x}_{10}^{\, 2}\, ,\\
& & \\
\hat{G} & = & {\textstyle\frac{1}{4}}a_{ij}d\omega^{i} \wedge d\omega^{j}\, ,\\
  \end{array}
\right.  
\end{equation}

\noindent
where the $\omega^{i}$s are 1-forms defined by

\begin{equation}
\omega^{i} = x^{(i)}dx^{5+(i)}-x^{5+(i)}dx^{(i)}\, ,
\end{equation}

\noindent
and where the constants $c_{i}$ and $a_{ij}$ (with $a_{ii}=0$) have to
satisfy the following two conditions:

\begin{equation}
a_{(i)j}a_{j(i)}= 4c^{2}_{i}+c_{j}c_{j}\, ,
\hspace{1cm}
a_{ij}c_{j}= -{\textstyle\frac{1}{8}}\eta^{ijklm}a_{jk}a_{lm}\, ,
\end{equation}

\noindent 
where the $\eta^{ijklm}$ symbol is one is all its indices take
different values and zero otherwise. It is convenient to label the
solutions of this family by the two vectors of constants

\begin{equation}
(c_{1},c_{2},c_{3},c_{4},c_{5})\, ,
\hspace{1cm}
(a_{12}, a_{13},a_{14},a_{15},a_{23},a_{24},a_{25},a_{34},a_{35},a_{45})\, .  
\end{equation}

The supersymmetry of some explicit G\"odel solutions was studied in
\cite{Harmark:2003ud} and some examples are given in the
table~\ref{tab:MGodelspaces}

\begin{table}[htbp]
  \centering
  \begin{tabular}{|c|c|c|c|}
\hline
& $(c_{i})$ & $(a_{ij})$ & \# \textrm{Susies} \\
\hline
(i) & $\lambda (1,1,0,0,0)$ & $\lambda (0,-2,-2,-2,-2,-2,-2,0,0,0)$ & 20 \\
\hline
(ii) & $\lambda (1,1,1,1,0)$ & $\lambda (4,0,0,-2,0,0,-2,4,-2,-2,-2)$ & 0 \\
\hline
(iii) & $\lambda (1,1,1,-1,0)$ & $\lambda (4,0,0,-2,0,0,-2,4,-2,-2,+2)$ & 20 \\
\hline
(iv) & $\lambda (1,0,0,0,0)$ & $\lambda (2,0,0,-2,0,0,0,2,0,0)$ & 8 \\
\hline
(v) & $\lambda (2,1,1,1,1)$ & $\lambda (-6,2,2,2,0,0,0,4,4,4)$ & 18 \\
\hline
  \end{tabular}
\caption{Some G\"odel solutions of $N=1,d=11$ SUGRA. The first example 
gives, after dimensional reduction, the maximally supersymmetric G\"odel 
solution of $N=1,d=5$ SUGRA Eq.~(\ref{eq:Goedel}) and, therefore, it also 
gives rise to the flacuum solution Eq.~(\ref{eq:flacuum}).}
\label{tab:MGodelspaces}
\end{table}

It is clear that the timelike reduction of any of these G\"odel
spacetimes gives a flacuum solutions of Euclidean type~IIA SUGRA with
flat metric and RR 4-form and 1-form given by

\begin{equation}
\left\{
  \begin{array}{rcl}
ds^{2} & = &  -d\vec{x}_{10}^{\, 2}\, ,\\
& & \\
G^{(2)} & = & c_{i}d\omega^{i}\, ,\\
& & \\
G^{(4)} & = & {\textstyle\frac{1}{4}}a_{ij}d\omega^{i} \wedge d\omega^{j}\, .\\
  \end{array}
\right.  
\end{equation}

It is important to understand how the vanishing of the energy-momentum
tensor takes place and why the dilaton is constant in this case. The
action of the Euclidean type~II SUGRA obtained by dimensional
reduction from standard, Lorentzian 11-dimensional SUGRA is given by

\begin{equation}
\label{eq:IIA_alpha}
\begin{array}{rcl}
S & = &{\displaystyle\int} d^{10}x\sqrt{|g|}
\biggl\{
e^{-2\phi}\left[
R - 4\left(\partial\phi\right)^{2}
-\textstyle{1\over 2\cdot 3!}\alpha_{11}H^{2}\right] 
\ +\  \textstyle{\textstyle\frac{1}{4}} \alpha_{11} G^{(2)\, 2} 
\ -\  \textstyle{1\over 2\cdot 4!} G^{(4)\, 2} \\
& & \\
& & 
\hspace{2cm}
-{\textstyle\frac{1}{144}} \frac{1}{\sqrt{|{g}|}}\
{\epsilon}\partial{C}^{(3)}\partial{C}^{(3)}{B}
\biggr\}
\end{array}
\end{equation}

This action differs from the type~IIA action by the occurrences of
$\alpha_{11}$'s, the definition of the various field-strengths staying
the same\footnote{Our conventions here are the same as in
  Ref.~\cite{Meessen:1998qm}.}. Observe that, to obtain this bosonic
action by a standard Wick rotation one has to consider both the RR
1-form and the NSNS 2-form pseudotensors.

The RR 1- and 3-form kinetic terms $G^{(2)\, 2}$ and $G^{(4)\, 2}$
have opposite signs when $\alpha_{11}=+1$ and the dilaton does not
couple directly to them in the string frame, but only through the
Ricci scalar. Thus, the exact cancellation between the respective
``energy-momentum'' tensors which makes the metric flat is enough to
render the dilaton constant. 

To end this section, let us observe that the BPST instanton
\cite{Belavin:fg} also provides an example of flacuum solution since
its energy-momentum tensor is identically zero because of the (anti-)
selfduality of its field strength on the round $S^{4}$. Thus, it is a
solution of the Euclidean Einstein-Yang-Mills theory with positive
cosmological constant proportional to the square inverse radius of the
sphere. The embedding of this solution in a supergravity theory is
more problematic because a $DS$-type gauged SUGRA would have to be
used. It should be possible to uplift the solution to Euclidean purely
gravitational $d=7$ solution with the $S^{7}$ metric
\cite{Trautman:1977im}


\subsection{$pp$-waves, lightlike reductions and flacuum solutions}

The fact that the $N=2,d=4$ flacuum solution originates in a timelike
and spacelike dimensional reduction of the KG6 H$pp$ wave suggests
that there must exist a relation between this kind of solutions and
$pp$-waves. 

$pp$-wave metrics can always be put in the  form

\begin{equation}
\hat{d}s^{2} = 2 du (dv + K du +A_{\underline{i}}  d x^{i})
+g_{\underline{i}\underline{j}}  dx^{i} d x^{j} \, .
\end{equation}

\noindent 
where all the functions in the metric are independent of $v$.  $K$ or
$A_{\underline{i}}$ can be removed by coordinate transformations that
preserve the above form of the metric, but we will keep both.
Furthermore, we are going to assume independence on the second
light-cone coordinate $u$

The components of the $\hat{d}$-dimensional Ricci tensor are

\begin{equation}
\hat{R}_{\underline{i}\underline{j}}=
R_{\underline{i}\underline{j}}\, , 
\hspace{1cm}
2 \hat{R}_{\underline{i}\underline{u}}=
\nabla^{\underline{j}}
F_{\underline{j}\underline{i}}\, ,
\hspace{1cm}
\hat{R}_{\underline{u}\underline{u}} =
\nabla_{\underline{i}}
\partial^{\underline{i}}K
-\textstyle{\frac{1}{4}}F^{2}\, ,
\end{equation}

\noindent 
where all the objects without hats are calculated with the
$d=(\hat{d}-2)$-dimensional Euclidean wavefront metric
$g_{\underline{i}\underline{j}}$ and where $F_{\underline{ij}}=
2\partial_{[\underline{i}}A_{\underline{j}]}$.  The Ricci scalar is
just $\hat{R}=R$ and, therefore, $\hat{d}$-dimensional vacuum Einstein
equations $\hat{G}_{\hat{\mu}\hat{\nu}}=0$ imply the $d$-dimensional
vacuum Einstein equations $G_{\underline{i}\underline{j}}=0$, the
vacuum Maxwell equation in the background metric
$\nabla_{\underline{j}} F^{\underline{j}\underline{i}}=0$ and the
scalar equation of motion $\nabla^{2}K=\frac{1}{4}F^{2}$.

This is similar to the result of a Kaluza-Klein reduction, except for
the fact that neither the scalar nor the vector field contribute to
the Euclidean ``energy-momentum'' tensor\footnote{Since either of them
  can be removed by reparametrizations, only one of them may
  contribute.} as in flacuum solutions. Actually, the flacuum solution
can be embedded in a purely gravitational $\hat{d}=6$ $pp$-wave with
$K\sim\lambda |\vec{x}_{4}|^{2}$. Another wide class of similar
solutions is given by 

\begin{equation}
F_{\underline{i}\underline{\sharp}}=2\nabla_{\underline{i}}H\, ,
\hspace{1cm}
F_{\underline{i}\underline{j}}=\epsilon_{ijk}
{\textstyle\frac{1}{\sqrt{|g|}}}\nabla_{\underline{k}}H
\hspace{1cm}
K=H^{2}\, ,
\hspace{1cm}
\nabla^{2}H=0\, .
\hspace{1cm}  
\end{equation}

It is possible to perform dimensional reductions in one lightlike
dimension \cite{Julia:1994bs}, but it does not very useful as the
resulting theories have singular metric. The above observation
suggests that lightlike reductions should be made in the two light-cone
coordinates to give a Euclidean $\hat{d}-2$-dimensional theory that
has essentially a metric an a vector field constrained to have
vanishing energy-momentum tensor and to satisfy the constraint
$F^{2}=4\nabla^{2}K$ for some scalar $K$. In $d=4$ it is possible to
define Euclidean theories in which the selfdual or anti-selfdual parts
of the spin connection are the only variables. It might be possible to
construct a theory in which only the selfdual or anti-selfdual part of
$F$ occurs and we can speculate that this kind of theory is the one
associated to lightlike dimensional reductions.


\section{Limiting Solutions}

Apart from the dimensional reduction from the KG6 solution, there is
another way of obtaining the G\"odel solution through a limiting
procedure which is very similar to the Penrose-G\"uven limit
\cite{kn:Pen6,Gueven:2000ru} that relates $aDS_{n}\times S^{m}$
supergravity vacua to Kowalski-Glikman-type H$pp$-wave solutions and
in general any supergravity solution to a plane wave solution with
equal or larger number of isometries and supersymmetries
\cite{Blau:2002dy,Blau:2002rg}.

Let us consider the NH limit of the extreme over-rotating BMPV BH
Eq.~(\ref{eq:overrotatingBMPV}). Its $\xi\rightarrow 0$ limit is well
defined and gives the vacuum with $aDS_{3}\times S^{2}$ geometry.  The
naive limit $\xi\rightarrow\infty$ is singular, though, but can be
made regular by using the following observation: under the field
rescalings 

\begin{equation}
\label{eq:rescalings}  
e^{a}{}_{\mu}\rightarrow \varepsilon^{-1}e^{a}{}_{\mu}\, ,
\hspace{1cm}
V\rightarrow \varepsilon^{-1}V\, ,
\end{equation}

\noindent
the action of $N=1,d=5$ SUGRA Eq.~(\ref{eq:N2d5SUGRAaction}) scales
homogeneously $S\rightarrow \varepsilon^{-3}S$. Then, we can combine
this rescaling that transforms solutions into solutions with the
reparametrization

\begin{equation}
\chi\rightarrow \varepsilon\chi\, ,
\hspace{.5cm} 
\theta\rightarrow \varepsilon\theta\, ,
\hspace{.5cm} 
t\rightarrow \varepsilon t +
R_{3}/2(\cosh\xi\phi -\sinh\xi\varphi)\, ,
\end{equation}

\noindent
and the gauge transformation of the vector field

\begin{equation}
\delta V = {\textstyle\frac{\sqrt{3}}{2}}R_{3} 
(\cosh\xi d\varphi -\sinh\xi d\phi)\, ,  
\end{equation}

\noindent
and take the double limit $\xi\rightarrow\infty$,
$\varepsilon\rightarrow 0$, whilst keeping the product
$\varepsilon\sinh\xi=1$. The result is the solution

\begin{equation}
\label{eq:Goedel2}
\left\{
  \begin{array}{rcl}
ds^{2} & = &  
[dt +R_{3}/4 (\chi^{2}d\phi -\theta^{2} d\varphi )]^{2} 
-(R_{3}/2)^{2}[d\chi^{2} +\chi^{2}d\phi^{2}
+d\theta^{2}+\theta^{2}d\varphi^{2}]\, ,\\
& & \\
V & = & \frac{\sqrt{3}}{4}R_{3}(\theta^{2}d\varphi +\chi^{2}d\phi)\, , 
  \end{array}
\right.
\end{equation}

\noindent
which is the G\"odel solution Eq.~(\ref{eq:Goedel}) in polar
coordinates with $\lambda=1/R_{3}$.

In this limit the number of supersymmetries and symmetries has been
preserved.  Observe that the gravitino supersymmetry rule
Eq.~(\ref{eq:N=2d=5susyrule}) $\delta_{\epsilon}\psi_{\mu}$ is
invariant under the above field rescalings. Actually, the arguments of
\cite{Blau:2002dy,Blau:2002rg} for the Penrose-G\"uven limit apply
equally to this kind of limit.

In the remaining of this section we are going to apply it to two
cases: the Euclidean Robinson-Bertotti solutions of $N=2_{\pm},d=4$
SUGRA Eqs.~(\ref{eq:H2S2-}),(\ref{eq:H2S2+}) and the $aDS_{5}\times
S^{5}$ solution of $d=10$ type~IIB SUGRA.


\subsection{The limit of the Euclidean Dyonic Robinson-Bertotti solutions}
 
The action and gravitino supersymmetry rule of $N=2,d=4$ SUGRA also
transform homogeneously under the rescalings Eq.~(\ref{eq:rescalings})
and we can use them to take a smooth limit of the Euclidean dyonic
Robinson-Bertotti solutions Eqs.~(\ref{eq:H2S2-}),(\ref{eq:H2S2+}),
exactly as in the previous case. The calculations are clearly
identical and the result is, not surprisingly the flacuum solution
Eq.~(\ref{eq:flacuum}).


\subsection{Boosted Hopf oxidation of $aDS_{5}\times S^{5}$}
\label{app-aDS5S5}

It is well known that $S^{5}$ can be seen as an $S^{1}$ fibration over
$\mathbb{CP}^{2}$.  In \cite{Pope:1999xg}, it was observed that the
$aDS_{2n+1}$ spacetimes can also be seen as a timelike $S^{1}$
fibration over a non-compact version of $\mathbb{CP}^{n}$, which the
authors called $\overline{\mathbb{CP}}^{n}$.  This construction was,
then, used to derive solution in the Euclidean IIA theory via timelike
T-duality. 

The relation of the NH limit of the extreme BMPV BH solutions to the
$N=(1,0), d=6$ vacuum $aDS_{3}\times S^{3}$, however, shows that in
general we before performing dimensional reduction or a T-duality
transformation one can mix the $S^{1}$ fibers of the sphere and the
$aDS$ spacetime by a rotation, if both fibers are spacelike, or by a
boost if one is spacelike and the other is timelike.

In the case of $aDS_{5}\times S^{5}$ the only mixing that we can do is
a boost on the timelike fiber in the $aDS_{5}$ and the spacelike fiber
for the $S^{5}$. After the boost, we have basically 2 options:
T-dualize over the new timelike direction, obtaining a generalization
of the solutions discussed in \cite{Pope:1999xg}, or T-dualize over
the new spacelike direction, which will lead to a solution similar to
the NH limit of the extreme over-rotating BMPV BH, and that will
contain the $aDS_{5}\times\mathbb{CP}^{2}\times\mathbb{R}$ solution as
a special limit.

Let us write the $aDS_{5}\times S^{5}$ solution as 

\begin{equation}
\left\{
\begin{array}{rcl}
ds_{IIB}^{2} & = & 
\left[ dt - B(\xi,\overline{\xi})\right)]^{2} \ -\ 
\left[dy - A(z,\overline{z})\right]^{2} \ -\
d\overline{\mathbb{CP}}^{2}_{[\xi]} \ -\
d\mathbb{CP}^{2}_{[z]} \; , \\
 & &  \\
G^{(5)} & = & 
4\ dt\wedge \omega_{\overline{\mathbb{CP}}^{2}} \, +\, 
4\ dy\wedge \omega_{\mathbb{CP}^{2}} \; ,\\
\end{array}
\right.
\end{equation}

\noindent
where we have introduced ($i,j=1,2$) the metrics

\begin{equation}
\begin{array}{rclrcl}
d\mathbb{CP}^{2}_{ [z]} & = & 
{\displaystyle\frac{dz^{i}d\bar{z}^{i}}{1+ |z|^{2}}\, -\,
\frac{\bar{z}^{i}dz^{i}\ z^{j}d\bar{z}^{j}}
{(1+|z|^{2})^{2}}}\, ,\hspace{1cm} &
A & = & 
{ i\over 2}{\displaystyle\frac{\bar{z}^{i}dz^{i}
-z^{i}d\bar{z}^{i}}{1+|z|^{2}}}\, , \\
 & & & & &  \\
d\overline{\mathbb{CP}}^{2}_{ [\xi ]} & = & 
{\displaystyle{d\xi^{i}d\bar{\xi}^{i}\over 1- |\xi |^{2}} \ +\
{\xi^{i}d\bar{\xi}^{i}\ \bar{\xi}^{j}d\xi^{j}\over 
\left( 1-|\xi |^{2}\right)^{2}}}\, ,&
B & = & {i\over 2}{\displaystyle{\xi^{i}d\bar{\xi}^{i} \ -\ 
\bar{\xi}^{i}d\xi^{i}\over 1-|\xi |^{2}}} \; ,\\
  \end{array}
\end{equation}

\noindent
$\omega_{\mathbb{CP}^{2}}$ and $\omega_{\mathbb{CP}^{2}}$ being their
corresponding volume 4-forms.

Boosting on the $(t,y)$-plane, with boost parameter $\beta$,
T-dualizing over $y$ and lifting the  solution up to M-theory, we find
the solution

\begin{equation}
\left\{
  \begin{array}{rcl}
ds^{2}_{11} & = & 
\left( dt -\cosh\beta B
-\sinh\beta A\right)^{2} \ -\
d\overline{\mathbb{CP}}^{2}_{ [\xi ]}\ 
-\ d\mathbb{CP}^{2}_{ [z]} \ -\ d\mathbb{R}^{2}_{[x]} \; ,\\
& &  \\
G & = & 
4 \cosh\beta\ \omega_{\mathbb{CP}^{2}} -
4 \sinh\beta\ \omega_{\overline{\mathbb{CP}}^{2}} 
-\left[ \sinh\beta dB +
  \cosh\beta dA \right]\wedge dx^{1}\wedge dx^{2} \; .\\
\end{array}
\right.
\end{equation} 

Now we can take the $\beta\rightarrow \infty$ limit using the fact
that the action for the bosonic fields of M-theory scales
homogeneously under the field rescalings

\begin{equation}
\label{eq:rescalings2}  
e^{a}{}_{\mu}\rightarrow \varepsilon^{-1}e^{a}{}_{\mu}\, ,
\hspace{1cm}
C_{\mu\nu\rho}\rightarrow \varepsilon^{-3}C_{\mu\nu\rho}\, .
\end{equation}

\noindent 
We perform the above rescaling and a change of coordinates

\begin{equation}
t\rightarrow\varepsilon t\, ,
\hspace{.5cm} 
\xi^{i}\rightarrow\varepsilon\xi^{i}\, ,
\hspace{.5cm} 
z^{i}\rightarrow \varepsilon z^{i}\, ,
\hspace{.5cm}
x^{i}\rightarrow \varepsilon x^{i}\, ,
\end{equation}

\noindent
and now take the double limit $\beta\rightarrow \infty$
$\varepsilon\rightarrow 0$ with $\sinh\beta\varepsilon=1$. It is
readily seen that the limiting solution is the $n=4$ G\"odel spacetime
that preserves 20 supersymmetries \cite{Harmark:2003ud}.


\section{Conclusions}
\label{sec-conclusions}

In this paper we have explored the relations between maximally
supersymmetric vacua of $d=4,5,6$ Euclidean and Lorentzian SUGRA
theories with 8 supercharges and we have found that the maximally
supersymmetric graviphoton background of
\cite{Seiberg:2003yz,Berkovits:2003kj} which we have called flacuum
solution is related to the maximally supersymmetric G\"odel solution
by timelike dimensional reduction, a situation that can be generalized
to higher-dimensional G\"odel solutions of M-theory, for instance, and
higher-dimensional flacuum solutions, of Euclidean type~II theory, for
instance.

We have studied other instances in which Euclidean solutions with
vanishing energy-momentum tensor appear, as in the (``double'')
lightlike dimensional reduction.

Apart from the relations between solutions via spacelike and timelike
dimensional reduction and oxidation we have found a limiting procedure
similar in may respects to the well-known Penrose-G\"uven limiting
procedure that, instead of giving plane-wave spacetimes gives, at
least in the examples considered, G\"odel solutions.  

The main interest in flacuum solutions, for the moment, seems to be
due to the fact that their superalgebras are small deformations of the
superalgebra that ``defines'' the SUGRA theory in question. All the
maximally supersymmetric vacua if a given SUGRA have, in fact, a
symmetry superalgebra that can be seen as a deformation of the SUGRA
superalgebra in which some central charges are activated and these,
together with some momenta, are given non-trivial commutation
relations between them and with the supercharges. These new
non-vanishing commutators and the central charges vanish for some
value of the deformation parameter.

For instance, the (Lorentzian) $N=2,d=4$ Poincar\'e superalgebra is
given by the only on-vanishing anticommutator (in a Majorana basis)

\begin{equation}
\{\mathcal{Q}_{(\alpha i)},\mathcal{Q}_{(\beta j)}\} =
-i\delta_{ij}(\mathcal{C}\gamma^{a})_{\alpha\beta}P_{a}
+i\mathcal{C}_{\alpha\beta}  \epsilon_{ij} Q
+(\mathcal{C}\gamma_{5})_{\alpha\beta}  
\epsilon_{ij} P\, ,
\end{equation}

\noindent
where $Q$ and $P$ are the central charges. All the bosonic charges
commute.  Now, let us consider different maximally supersymmetric
vacua of this theory. Minkowski spacetime has the same superalgebra,
with vanishing central charges. The superalgebra of the
Robinson-Bertotti solution ($aDS_{2}\times S^{2}$) has the same
anticommutator but, now, the six bosonic charges generate an
$SO(2,1)\times SO(3)$ Lie algebra

\begin{equation}
  \begin{array}{rclrclrcl}
[Q,P_{1}] & = & \frac{1}{R_{2}}P_{3}\, ,\hspace{.5cm} &
[Q,P_{0}] & = & \frac{1}{R_{2}}P_{1}\, ,\hspace{.5cm} &
[P_{1},P_{0}] & = & -\frac{1}{R_{2}}Q\, ,\\
& & & & & & & & \\
\left[P,P_{2}\right] & = & \frac{1}{R_{2}}P_{3}\, ,\hspace{.5cm} &
[P,P_{3}] & = & -\frac{1}{R_{2}}P_{2}\, ,\hspace{.5cm} &
[P_{2},P_{3}] & = & \frac{1}{R_{2}}P\, ,\\
  \end{array}
\end{equation}

\noindent
that become again Abelian when the $aDS_{2}$ and $S^{2}$ radius
$R_{2}$ goes to infinity. The commutators of the supercharges and
these bosonic generators also become nontrivial and have a similar
behavior.

Another examples is provided by the KG4 H$pp$-wave solution
\cite{Kowalski-Glikman:im}. The supercharges of its symmetry
superalgebra also have the same anticommutator as above, but now the
six bosonic charges generate the Heisenberg algebra with non-vanishing
commutators

\begin{equation}
  \begin{array}{rclrclrcl}
[Q,P_{1}] & = & -\frac{\lambda}{4} P_{v}\, ,\hspace{.5cm} &
[P,P_{2}] & = & -\frac{\lambda}{4} P_{v}\, ,\hspace{.5cm} &
[P_{u},P_{1}] & = & \lambda Q\, ,\\
& & & & & & & & \\
\left[P_{u},P_{2}\right] & = & \lambda P\, ,\hspace{.5cm} &
[P_{u},Q] & = & -\frac{\lambda}{4} P_{1}\, ,\hspace{.5cm} &
[P_{u},P] & = & -\frac{\lambda}{4} P_{2}\, ,\\
  \end{array}
\end{equation}

\noindent
that also becomes Abelian in the $\lambda\rightarrow 0$ limit.

The maximally supersymmetric $aDS_{n}\times S^{m}$ vacua of 11- and
10-dimensional supergravities can also be seen as deformations of the
Poincar\'e superalgebra and the same is valid for H$pp$-wave
backgrounds. 

It is natural to expect that quantum theories defined in these
backgrounds satisfy some non-commutative generalization of the
Heisenberg algebra. 


\section*{Acknowledgements}

We thank E.~Lozano-Tellechea for interesting conversations.
T.O.~would like to thank M.M.~Fern\'andez for her support.  The work
of T.O.~has been supported in part by the Spanish grant BFM2003-01090.

\appendix

\section{The detailed dimensional reductions}
\label{app-details}

The dimensional reduction of the Einstein-Hilbert term is identical in
the two cases $N=(1,0),\hat{d}=6$ and $N=1,\hat{d}=5$ SUGRA. Further,
in both the timelike and spacelike cases we can make the
\textit{Ansatz}

\begin{equation}
\label{eq:standardVielbeinAnsatz}
\left( \hat{e}_{\hat{\mu}}{}^{\hat{a}} \right) = 
\left(
\begin{array}{cc}
e_{\mu}{}^{a} & kA_{\mu} \\
&\\
0             & k        \\
\end{array}
\right)
\, , 
\hspace{1cm}
\left(\hat{e}_{\hat{a}}{}^{\hat{\mu}} \right) =
\left(
\begin{array}{cc}
e_{a}{}^{\mu} & -A_{a}   \\
& \\
0             &  k^{-1}  \\
\end{array}
\right)\, ,
\label{eq:KKbasis}
\end{equation}

\noindent 
where $A_{a}= e_{a}{}^{\mu} A_{\mu}$. All $d$-dimensional fields with
Lorentz indices will be assumed to have been contracted with the
$d$-dimensional Vielbeins. 

\noindent 
This \textit{Ansatz} gives the following non-vanishing components of
the Ricci rotation coefficients

\begin{equation}
\hat{\Omega}_{abc}=\Omega_{abc}\, ,
\hspace{.6cm}  
\hat{\Omega}_{ab\sharp}={\textstyle\frac{1}{2}}\alpha_{\hat{d}}  F_{ab}\, ,
\hspace{.6cm}  
\hat{\Omega}_{a\sharp\sharp}={\textstyle\frac{1}{2}} \alpha_{\hat{d}}
\partial_{a}\log{k}\, ,
\end{equation}

\noindent
and of the the spin connection $\hat{\omega}_{\hat{a}\hat{b}\hat{c}}$

\begin{equation}
\label{eq:standardspinconnectionreduction}
\begin{array}{rclrcl}
\hat{\omega}_{abc} & = & \omega_{abc}\, , &
\hat{\omega}_{ab\sharp} & = & -\frac{1}{2}\alpha_{\hat{d}} k F_{ab}\, ,
\\
& & & & & \\
\hat{\omega}_{\sharp bc} & = & \frac{1}{2} \alpha_{\hat{d}} k F_{bc}\, ,
\hspace{1.5cm}& 
\hat{\omega}_{\sharp b \sharp} & = & \alpha \partial_{b} \log{k}\, ,\\
\end{array}
\end{equation}

\noindent 
where

\begin{equation}
\label{eq:KKfieldstrength}
F_{ab} = e_{a}{}^{\mu} e_{b}{}^{\nu} F_{\mu\nu}\, ,
\hspace{1cm}
F_{\mu\nu} = 2\partial_{[\mu}A_{\nu]}\, ,
\end{equation}

\noindent
(which we will also denote by $F(A)$ in presence of other vector
fields) is the KK vector field strength. The standard procedure gives
the KK-frame action

\begin{equation}
\label{eq:KKactionKKframe}
\hat{S} = {\displaystyle\int}
 d^{\hat{d}}\hat{x}\ 
\sqrt{|\hat{g}|}\, \hat{R}
={\displaystyle\int} d^{\hat{d}-1}x \sqrt{|g|}\ k\left[R 
+{\textstyle\frac{1}{4}}\alpha_{\hat{d}} k^{2}F^{2}\right]\, .
\end{equation}

\noindent
The sign of the Maxwell term in the timelike case is unconventional
because we would get the opposite sign by Wick-rotating the standard
Lorentzian Maxwell action. However, it cannot be deemed as ``wrong''
since there is no propagation, no concepts of motion or energy in
Euclidean space. On the other hand, if we Wick-rotated a pseudovector
field we would get precisely the above sign.


\subsection{$N=(1,0),\hat{d}=6$ SUGRA}

In the reduction of the action of $N=(1,0),\hat{d}=6$ SUGRA
Eq.~(\ref{eq:d6action}) to $d=5$ we have to take into account the
anti-selfduality constraint, which can be imposed on the action
\textit{after} dimensional reduction, just as in the reduction of
$N=2B,d=10$ supergravity on a circle \cite{Meessen:1998qm}.
Thus, we first perform the naive dimensional reduction of
Eq.~(\ref{eq:d6action}).

We can use our previous results for the reduction of the
Einstein-Hilbert term and we only need to reduce the 2-form kinetic
term. The reduction of $\hat{H}^{-}$ gives a 3- and a 2-form field
strengths

\begin{equation}
  \begin{array}{rclrcl}
H_{abc} & = & \hat{H}^{-}_{abc}\, ,
\hspace{1cm}&
H & = & 3 \left[\partial B -\frac{1}{2}AF(B) -\frac{1}{2}BF(A)\right]\, , \\
& & & & & \\
F_{ab}(B) & = & k^{-1}\hat{H}^{-}_{ab\sharp}\, ,&
F(B) & = & 2\partial B\, ,\\
\end{array}
\end{equation}

\noindent
where

\begin{equation}
\hat{B}_{\mu\nu} =B_{\mu\nu}-A_{[\mu}B_{\nu]}\, ,
\hspace{1cm}
B_{\mu}=\hat{B}_{\mu\underline{\sharp}}\, .  
\end{equation}

The anti-selfduality constraint becomes

\begin{equation}
\label{eq:reducedantiselfdualityconstraint}
{}^{\star}H=-k^{-1}F(B)\,\,\, \Leftrightarrow\,\,\,
H=\alpha_{6}k^{-1}{}^{\star}F(B)\, .  
\end{equation}

We immediately get

\begin{equation}
S=\int d^{5}x\sqrt{|g|}\, k\,[R 
+{\textstyle\frac{1}{4}}\alpha_{6}k^{2}F^{2}(A)
+{\textstyle\frac{1}{8}}\alpha_{6}k^{-2}F^{2}(B)
+{\textstyle\frac{1}{24}}H^{2}]\, .
\end{equation}

As in Ref.~\cite{Meessen:1998qm}, we Poincar\'e-dualize the 2-form
into a third vector field $C_{\mu}$ and then we identify it with
$B_{\mu}$ in the action, that is left with the metric and the
unconstrained vector fields $A_{\mu}$ and $B_{\mu}$ and takes the form

\begin{equation}
\label{eq:N=2d=5action}
S=\int d^{5}x\sqrt{|g|}\, k\,[R 
+{\textstyle\frac{1}{4}}\alpha_{6}k^{2}F^{2}(A)
+{\textstyle\frac{1}{4}}\alpha_{6}k^{-2}F^{2}(B)
-\alpha_{6}{\textstyle\frac{\epsilon}{8\sqrt{|g|}}}k^{-1}F(A)F(B)B]\, .   
\end{equation}

The truncation to pure supergravity involves setting to zero the
scalar $k=1$ consistently, i.e.~in such a way that its equation of
motion is always satisfied. The $k$ equation of motion with $k=1$
(upon use of Einstein's equation) implies the constraint

\begin{equation}
\label{eq:firstconstraint}
F^{2}(B)=2F^{2}(A)\, .  
\end{equation}

We also have to set to zero the gaugino $\lambda$ and the matter
vector field $W_{\mu}$, which must be a combination of $A_{\mu}$ and
$B_{\mu}$ related to $\lambda$ by supersymmetry. Thus, to identify
$W_{\mu}$, we have to analyze the reduction and truncation of the
gravitino supersymmetry transformation rule Eq.~(\ref{eq:d6susy}).

The 6- and 5-dimensional gamma matrices are related by

\begin{equation}
  \begin{array}{rcl}
\hat{\gamma}^{a} & = & \gamma^{a}\otimes \sigma^{1}\, ,\\
& & \\
\hat{\gamma}^{\sharp} & = & \mathbb{I}\otimes 
\alpha_{6}^{1/2}\sigma^{2}\equiv 
\hat{\gamma}^{\sharp}\, ,\\
& & \\
\hat{\gamma}_{7} & = & \hat{\gamma_{0}}\cdots\hat{\gamma}_{5}
=\mathbb{I}\otimes \sigma^{3}\, ,\\
  \end{array}
\end{equation}

\noindent 
where the $\gamma^{a}$s Lorentzian or Euclidean 5-dimensional gamma
matrices satisfying $\gamma_{0}\cdots\gamma_{4}=\mathbb{I}$ and
$\gamma_{1}\cdots \gamma_{5}=i\mathbb{I}$. 

Using this relation, the decompositions of the spin connection and
3-form field strength, the anti-selfduality constraint
Eq.~(\ref{eq:reducedantiselfdualityconstraint}), the chirality of
$\hat{\epsilon}^{+}$ and assuming that the supersymmetry parameter
$\hat{\epsilon}^{+}$ is independent of $x^{\sharp}$ and
setting $k=1$, we find

\begin{equation}
  \begin{array}{rcl}
\delta_{\hat{\epsilon}^{+}}\hat{\psi}^{+}_{a} & = & 
\left\{\nabla_{a} -\frac{i}{4}\alpha_{6}^{3/2}\not\! F_{a}(A)
+\frac{i}{8\sqrt{2}}\alpha_{6}^{1/2}
\not\! F(B)\gamma_{a}\right\}\epsilon\, ,\\
& & \\
\delta_{\hat{\epsilon}^{+}}\hat{\psi}^{+}_{\sharp} & = & 
\frac{-1}{8\sqrt{2}}\left\{\sqrt{2}\alpha_{6}\not\! F(A) 
+\not\! F(B) \right\}\epsilon \, ,\\
  \end{array}
\end{equation}

\noindent
where $\frac{1}{2}(1+\sigma^{3})\hat{\epsilon}^{+}=\epsilon$.

$\hat{\psi}^{+}_{a}$ is to be identified with the 5-dimensional
gravitino $\frac{1}{2}(1+\sigma^{3})\hat{\psi}^{+}_{a}=\psi_{a}$.
$\hat{\psi}^{+}_{\sharp}$ only transforms into vectors and it is to be
identified with the gaugino
$\frac{1}{2}(1+\sigma^{3})\hat{\psi}^{+}_{\sharp}\equiv \lambda$.
This, in turn, implies that the combination of vector field strengths
in the r.h.s.~is proportional to $F(W)$ (for $k=1$). The field
strengths of $V$ and $W$ must be an $SO(2)$ rotation of those of $A$
and $B$ to preserve the canonical normalization and diagonal form of
the kinetic terms in the action. The right combination is that in
which $\lambda$ only transforms into $W$ and not into $V$:

\begin{equation}
  \begin{array}{rcl}
F(V) & \equiv & 
{\textstyle\frac{1}{\sqrt{3}}}F(A) 
-\sqrt{\textstyle\frac{3}{2}}\alpha_{6}F(B)\, ,\\
& & \\
F(W) & \equiv &
\sqrt{\textstyle\frac{2}{3}}\alpha_{6}F(A)
+{\textstyle\frac{1}{\sqrt{3}}} F(B)\, .
\end{array}
\end{equation}

Setting $W_{\mu}=0$ (to make consistent $\lambda=0$) implies the
constraint

\begin{equation}
\sqrt{2}\alpha_{6}F(A) =-F(B)\, ,  
\end{equation}

\noindent
which implies the constraint Eq.~(\ref{eq:firstconstraint}).

Therefore, we can truncate the vector multiplet setting consistently
$k=1,\lambda=0,W_{\mu}=0$. The result is the action of pure Lorentzian
or Euclidean $N=1,d=5$ SUGRA Eq.~(\ref{eq:N2d5SUGRAaction}) and its
gravitino supersymmetry transformation law
Eq.~(\ref{eq:N=2d=5susyrule}).


\subsection{$N=1,\hat{d}=5$ SUGRA}

The reduction of the Einstein-Hilbert term in the action
Eq.~(\ref{eq:N2d5SUGRAaction}) goes as in the previous case. The
reduction of the Maxwell field $\hat{V}_{\hat{\mu}}$ gives rise to a
$d$-dimensional vector field that we denote by $B_{\mu}$ and a scalar
$l$:

\begin{equation}
\hat{V}_{\underline{\sharp}}  =  l\, ,
\hspace{1cm}
\hat{V}_{\mu} =  B_{\mu}+lA_{\mu}\, .
\end{equation}

As for the field strength, we define $F_{\mu\nu}(B)$ and the
combination $\mathcal{G}_{ab}=\hat{F}_{ab}$

\begin{equation}
F_{\mu\nu}(B)=2\partial_{[\mu}B_{\nu]}\, ,
\hspace{1cm}
\mathcal{G}=F(B)+lF(A)\, ,
\end{equation}

\noindent 
where $F(A)$ is the KK vector field strength.  Taking into account

\begin{equation}
\hat{F}_{a\sharp}=k^{-1}\partial_{a}l\,    
\end{equation}

\noindent
we find that the reduction of the Einstein-Hilbert and Maxwell terms
of the action Eq.~(\ref{eq:N2d5SUGRAaction}) gives

\begin{equation}
\label{eq:reducedEMaction}
\begin{array}{c}
{\displaystyle\int} d^{4}x \sqrt{|g|}\ k\left[R 
+{\textstyle\frac{1}{2}}\alpha_{5}\alpha_{6} k^{-2}(\partial l)^{2}
+{\textstyle\frac{1}{4}}\alpha_{5} k^{2}F^{2}(A)
+{\textstyle\frac{1}{4}}\alpha_{6}\mathcal{G}^{\, 2}
\right]\, .\\
\end{array}
\end{equation}

The reduction of the Chern-Simons term is identical for both the
timelike and spacelike cases, using
$\hat{\epsilon}^{abcd\sharp}=\epsilon^{abcd}$, and gives after
integration by parts the $d=4$ action

\begin{equation}
\label{eq:N2d5SUGRAactionreduced}
\begin{array}{rcl}
S  & = &
{\displaystyle\int} d^{4}x\sqrt{|g|}\, k\, \biggl\{R 
+{\textstyle\frac{1}{2}}\alpha_{5}\alpha_{6} k^{-2}(\partial l)^{2}
+{\textstyle\frac{1}{4}}\alpha_{5} k^{2}F^{2}(A)
+{\textstyle\frac{1}{4}}\alpha_{6}\mathcal{G}^{\, 2} \\
& & \\
& & 
\hspace{2cm}
\left.
-{\textstyle\frac{1}{4\sqrt{3}}}\alpha_{6}k^{-1}l
{\textstyle\frac{\epsilon}{\sqrt{|g|}}}
[\mathcal{G}+2A\partial l]^{2}\right\}\, .\\
\end{array}
\end{equation}

This theory is pure (Euclidean $\alpha_{6}\alpha_{5}=-1$ or Lorentzian
$\alpha_{6}\alpha_{5}=+1$) $N=2,d=4$ SUGRA theory coupled to a
``matter'' vector supermultiplet that contains a vector and the two
scalars $k$ and $l$ \cite{Chamseddine:1980sp}.  The matter and
supergravity vector fields are combinations of the two vectors $A,V$.
To identify the mater vector field we can use the fact that
eliminating a matter supermultiplet is always a consistent truncation.
The equations of motion of $k$ and $l$ after setting $k=1$ and $l=0$
give the constraints

\begin{equation}
\left\{
  \begin{array}{rcl}
3\alpha_{5} F^{2}(A) +\alpha_{6}F^{2}(B) & = & 0\, ,\\
& & \\
\left[ \sqrt{3}F(A)-{}^{\star}F(B)\right]F(B) & = & 0\, .\\
  \end{array}
\right.
\end{equation}

\noindent
The second constraint is solved by 

\begin{equation}
\label{eq:truncationcondition}
\sqrt{3}F(A)-{}^{\star}F(B)=0, 
\end{equation}

\noindent
which also solves the first\footnote{Observe that we have
  $({}^{\star}F)^{2}=-\alpha_{6}\alpha_{5} F^{2}$ and
  ${}^{\star\star}F=-\alpha_{6}\alpha_{5}F$ because
  $\alpha_{5}\alpha_{6}=+1$ corresponds to Lorentzian signature and
  $\alpha_{5}\alpha_{6}=-1$ to Euclidean signature.}.  This implies
that the matter vector field that we denote by $W_{\mu}$ is
proportional to the combination $\sqrt{3}F(B)-{}^{\star}F(B)$ or to
its Hodge dual, at least for the $k=1,\,\,l=0$ case. The orthogonal
combination will be the field strength of the supergravity vector that
we will denote by $V_{\mu}$.

To fully identify the supergravity and matter vector field strengths
we have to check the consistency of the truncation from the
supersymmetric point of view, that is, in the gravitino supersymmetry
transformation rule Eq.~(\ref{eq:N=2d=5susyrule}) (adding hats
everywhere).  Assuming that the supersymmetry parameter
$\hat{\epsilon}$ is independent of the internal coordinate so it can
be identified with the 4-dimensional one $\epsilon$ we get and setting
from now on $k=1$ and $l=0$

\begin{equation}
  \begin{array}{rcl}
\delta_{\epsilon}\hat{\psi}_{a} & = & 
\left\{
\nabla_{a} -\frac{1}{4}\alpha_{5}\not\! F_{a}(A)\gamma^{\sharp}
-\frac{i}{8\sqrt{3}}\alpha_{6}^{3/2} 
[\not\! F(B) \gamma_{a} +2\not\! F_{a}(B)]
\right\} \epsilon\, ,\\ 
& & \\
\delta_{\epsilon}\hat{\psi}_{\sharp} & = &
-\frac{1}{8\sqrt{3}}\alpha_{5}
\left\{ \sqrt{3}\not\! F(A)+ \not\! F(B) 
\gamma^{\sharp} \right\} \epsilon\, ,
  \end{array}
\end{equation}

\noindent
where $\gamma^{a}=\hat{\gamma}^{a}$ with $a=0,1,2,3$ in the Lorentzian
case $\alpha_{5}\alpha_{6}=+1$,  and $a=1,2,3,4$ in the Euclidean case
$\alpha_{5}\alpha_{6}=-1$, and $\gamma^{\sharp}=\hat{\gamma}^{\sharp}$
is a matrix proportional to the 4-dimensional $\gamma_{5}$. It can be
shown that

\begin{equation}
  \begin{array}{rcl}
\not\!F_{a}\gamma^{\sharp} & = & 
-i\alpha_{6}^{1/2}
\left(\frac{1}{2} {}^{\star}\!\!\not\! F\gamma_{a} 
-{}^{\star}\!\!\not\! F_{a}\right)\, ,\\
& & \\
\not\! F \gamma^{\sharp} & = & 
i\alpha_{6}^{1/2}\, {}^{\star}\!\! \not\! F\, ,\\
\end{array}
\end{equation}

\noindent
and using these identities we can rewrite the gravitino supersymmetry
transformation rule as follows:

\begin{equation}
  \begin{array}{rcl}
\delta_{\epsilon}\hat{\psi}_{a} & = & 
\left\{
\nabla_{a} -\frac{i}{8}\alpha_{6}^{3/2}
\left[-\alpha_{5}\alpha_{6} {}^{\star}\!\!\not\!F(A) 
+\frac{1}{\sqrt{3}}\not\!F(B)\right]\gamma_{a}
+\frac{1}{4}\alpha_{6}^{3/2} 
\left[-\alpha_{5}\alpha_{6} {}^{\star}\!\!\not\!F_{a}(A) 
-\frac{1}{\sqrt{3}}\not\!F_{a}(B)\right]
\right\} \epsilon\, ,\\ 
& & \\
\delta_{\epsilon}\hat{\psi}_{\sharp} & = &
-\frac{1}{8}\alpha_{5}\left[\not\! F(A) 
-\frac{1}{\sqrt{3}}{}^{\star}\!\!\not\! F(B) \right] \epsilon\, ,
  \end{array}
\end{equation}

The components $\hat{\psi}_{a}$ clearly become the 4-dimensional
gravitino $\psi_{a}$.  The internal component of the gravitino
$\hat{\psi}_{\sharp}$ only transforms into a combination vector field
strengths and, thus, it can be identified with the gaugino $\lambda$
of the matter vector multiplet. The combination of vector field
strengths is proportional to the matter vector field strength $F(W)$
(for $k=1$ and $l=0$ only), in agreement with the constraint
Eq.~(\ref{eq:truncationcondition}) found in the truncation of the
action.  The normalization factor and the definition of the
supergravity vector field strength $F(V)$ are found by imposing the
diagonalization and correct normalization of the energy-momentum
tensors of the vectors in the Einstein equation: the r.h.s.~of the
Einstein equation contains

\begin{equation}
-{\textstyle\frac{1}{2}}\alpha_{5} T_{\mu\nu}(A)
-{\textstyle\frac{1}{2}}\alpha_{6}T_{\mu\nu}(B)\, ,
\end{equation}

\noindent
where $T_{\mu\nu}$, the energy-momentum tensor of a vector field in
four Euclidean or Lorentzian dimensions, is usually written in the
form

\begin{equation}
T_{\mu\nu}= 
F_{\mu}{}^{\rho}F_{\nu\rho}
-{\textstyle\frac{1}{2}}g_{\mu\nu}F^{2}\, ,
\end{equation}

\noindent
but can be rewritten in the more useful form

\begin{equation}
\label{eq:vectorenergymomentumtensor}
T_{\mu\nu}= {\textstyle\frac{1}{2}}
\left[F_{\mu}{}^{\rho}F_{\nu\rho}
+\alpha_{5}\alpha_{6} 
{}^{\star}F_{\mu}{}^{\rho} {}^{\star}F_{\nu\rho}\right]\, .
\end{equation}

This still leaves some ambiguities in the identification of the
supergravity vector field, but they are irrelevant when $W=0$ and lead
to a unique relation

\begin{equation}
F(V) = 2\alpha_{5}\alpha_{6} {}^{\star}F(A)\, .
\end{equation}

Now we can eliminate consistently the gaugino $\lambda$, the matter
vector field $W_{\mu}$ and scalars $k$ and $l$, getting the bosonic
action\footnote{The bosonic action cannot be found by performing the
  above field redefinitions and has to bes constructed so as to
  reproduce the truncated equations of motion.} and gravitino
supersymmetry transformation rule which are given in
Eq.~(\ref{eq:N2d4SUGRA}).


\end{document}